\documentclass[10pt,journal,compsoc,tikz]{IEEEtran}
\ifCLASSOPTIONcompsoc
  \usepackage[nocompress]{cite}
\else
  \usepackage{cite}
\fi

\ifCLASSINFOpdf
   \usepackage[pdftex]{graphicx}
   \graphicspath{{../pdf/}{../jpeg/}}
   \DeclareGraphicsExtensions{.pdf,.jpeg,.png}
\else
  \usepackage{graphicx}
  \graphicspath{{./images/}}
  \DeclareGraphicsExtensions{.eps}
\fi
\usepackage{graphicx}
\usepackage{comment}
\usepackage{multirow} 
\usepackage[table,xcdraw]{xcolor}
\usepackage{caption}
\usepackage{subcaption}
\usepackage{amsmath}
\usepackage{array}
\usepackage{tikz}

\usepackage[finalnew]{trackchanges}

\begin{document}
%\title{How to Overlay Quantitative Data On Networks}
\title{Overlaying Quantitative Measurement on Networks: An Evaluation of Three Positioning and Nine Visual Marker Techniques}

%: Ranking Three Positioning and Nine Visual Markers

\author{Guohao~Zhang,~\IEEEmembership{Student Member,~IEEE,}
        Alexander~P.~Auchus,
        Peter~Kochunov,
        Niklas~Elmqvist,~\textit{Senior~Member,~IEEE,}
        and~Jian~Chen,~\textit{Member,~IEEE}% <-this % stops a space
\IEEEcompsocitemizethanks{\IEEEcompsocthanksitem G. Zhang is with University of Maryland, Baltimore County. E-mail: guohaozhang@umbc.edu.
\IEEEcompsocthanksitem AP. Auchus is with the Neurology Department at University of Mississippi Medical Center. E-mail: aauchus@umc.edu.
\IEEEcompsocthanksitem P. Kochunov is with Maryland Psychiatric Research Center. E-mail: pkochunov@mprc.umaryland.edu.
\IEEEcompsocthanksitem N. Elmqvist is with University of Maryland, College Park. E-mail: elm@umd.edu.
\IEEEcompsocthanksitem J. Chen is with the Department of Computer Science and Engineering, The Ohio State University, Columbus, OH. 43210.\protect\\E-mail: chen.8028@osu.edu.
}
% <-this % stops an unwanted space
\thanks{Manuscript received October 9, 2016; revised August 26, 2015.}}

\markboth{Journal of \LaTeX\ Class Files,~Vol.~14, No.~8, August~2015}%
{Zhang \MakeLowercase{\textit{et al.}}: Overlaying Quantitative Measurement On Brain Networks: An Evaluation of 
Three Positioning and Nine Visual Marker Techniques}

\IEEEtitleabstractindextext{%
\begin{abstract}
%We conduct evaluation on the effectiveness of encoding node attributes using nine visual variables in brain functional network visualizations, combined with three network layout conditions. Our experiment studies markers constructed from nine visual variables (length, shape, density, slope, angle, size, texture,  lightness, and  hue) for encoding network centrality in three layout conditions (ring, projection, and matrix). We describe a task topology for network visualization that suits the emerging fMRI network analyses. Four tasks are then selected for the study (change-detection, neighbor-hub, lobe-hub and hemisphere-hub). Our results showed that the encoding rank has changed due to less emphasize on differentiating small difference in our experiment.  Our results also indicated that the ring layout led to most accurate answers and lest completion time overall compared to projection and  matrix views due to a linear arrangement of node and easy tracing.
%using the most recent visual marker  design methods. 
%using examples from brain connectivity studies 
We report results from an experiment on ranking visual markers and node positioning  techniques for network visualizations. Inspired by prior ranking studies, we rethink the ranking when the dataset size increases and when the markers are distributed in space. Centrality indices are visualized as node attributes. Our experiment studies nine visual markers and three positioning methods. Our results suggest that direct encoding of quantities improves accuracy by about $20\%$ compared to previous results. Of the three positioning techniques, circular was always in the top group, and matrix and projection switch orders depending on two factors: whether or not the tasks demand symmetry, or the nodes are within closely proximity. Among the most interesting results of ranking the visual markers for comparison tasks are that hue and area fall into the top groups for nearly all multi-scale comparison tasks; Shape (ordered by curvature) is perhaps not as scalable as we have  thought  and can support more accurate answers only when two quantities are compared; Lightness and slope are least accurate for quantitative comparisons regardless of scale of the comparison tasks. Our experiment is among the first to acquire a complete picture of ranking visual markers in different scales for comparison tasks.
\end{abstract}

}

\maketitle

\IEEEdisplaynontitleabstractindextext
\IEEEpeerreviewmaketitle

\IEEEraisesectionheading{\section{Introduction}\label{sec:introduction}}

\begin{table*}[h!]
  \centering
  \begin{tabular}{ l | l l l }
    \hline
    \begin{minipage}{\textwidth}
     \includegraphics[width=\linewidth]{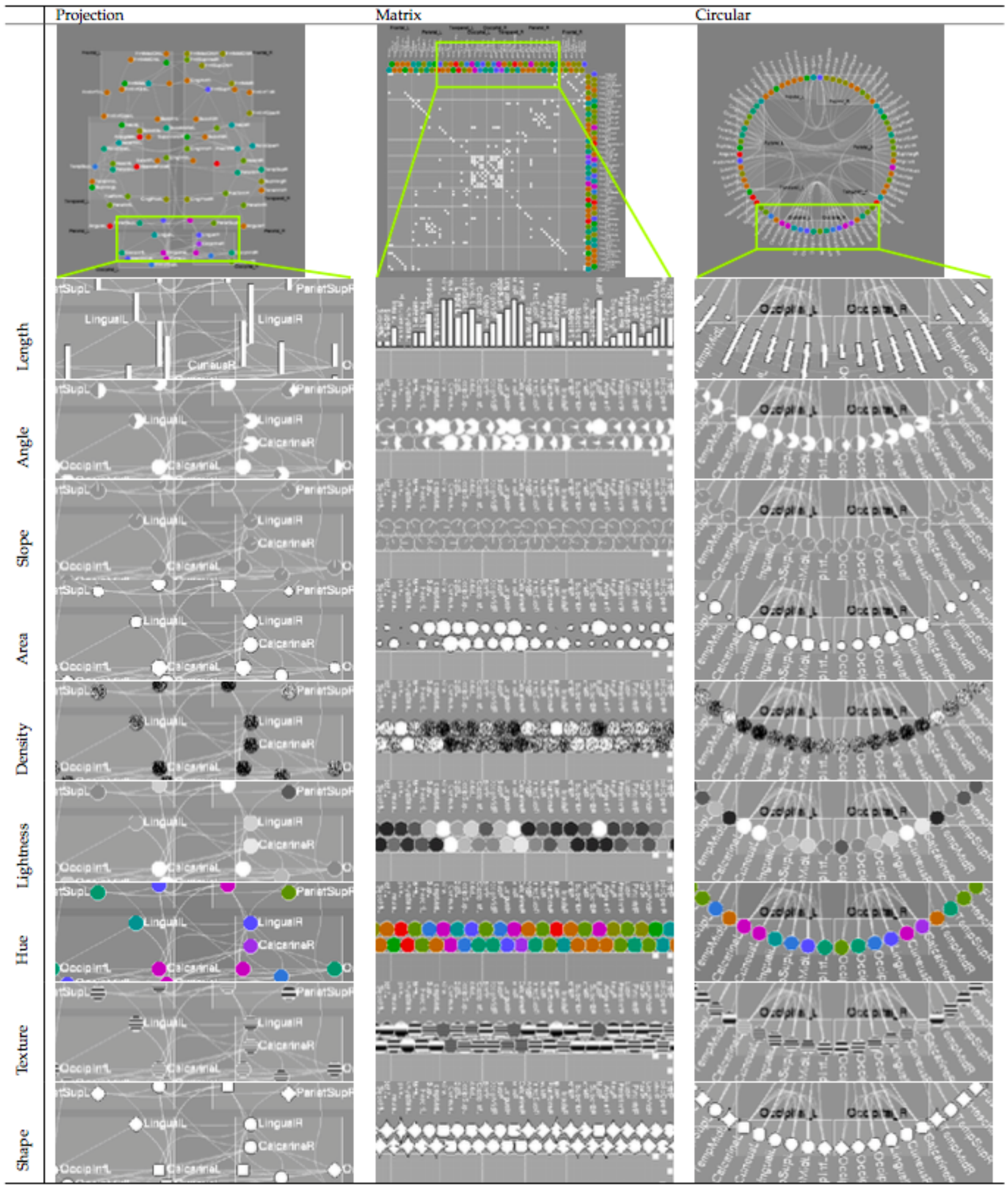}
    \end{minipage}
            \\ \hline   
            \end{tabular}
  \caption{Two-dimensional network visualization compared in this work.
	        They includes combinations of three layout (projection, circular,
		and matrix) and nine visual markers for showing the 
		centralities computed from a 74-node network.
		%The bottom right subfigure shows the nine visual markers, from left to right and 
		The nine marker types from top to bottom: length, angle, slope, area, 
		density, lightness, hue, texture, and shape. 
A pilot study also revealed no statistically significant differences between white- and gray-background color schemes on 
task completion time and accuracy.}
  \label{tab:taxonomy}
\end{table*}

\IEEEPARstart{M}{any}
\remove{
Functional magnetic resonance imaging (fMRI)  is  a non-invasive method in  which  
blood-oxygen-level-dependent (BOLD) signals are collected  over  time  to detect 
neuronal activation{~\cite{ogawa1990brain}}. The correlations between each  pair of temporal sequences 
are computed on brain regions{~\cite{Allen11112012}} to describe the functional connectivity, where these regions}
real-world datasets can  be  described as  networks:  entities are represented as nodes~\cite{zalesky2010network} 
and  
the presence  or  absence   of  an  edge  linking two  nodes describes the presence or  absence   of  direct  connectivity,  
often  defined by  thresholding correlation strength between nodes~\cite{rubinov2010complex}.   
Interpreting complex networks of relationships is, however, challenging, and that fact makes  visualization an  indispensable  tool  in  practice  for  deciphering unexpected or revealing interesting results~\cite{margulies2013visualizing}.

\remove{The integration of
fMRI in neurobiological studies has enabled the examination of the brain systems 
underlying many behavioral deficits manifested in neurodegeneration diseases 
such as schizophrenia and bipolar disorders{~\cite{shergill2000mapping}}.} 
%\remove{~\cite{saad2012trouble}.}

%Great strides have been made to visualize fMRI methods, e.g., 3D methods to show the
%~\cite{bullmore2012economy, zuo2012network}.
%A node with high \textit{betweenness centrality} controls information 
%flow because it is at the intersection of many short paths. 
\remove{Since schizophrenia patients tends to lose
connectivity among brain regions in the lost of the functional integration for adaptive behavior, changing
in hubs are associated with schizophrenia{~\cite{cheng2015nodal}}. }
\remove{This centrality has been used across the entire brain networks to compare hallucinations
between patient and normal control groups{~\cite{chen2015abnormal}}.  
Often the centrality of hubs is also lower in schizophrenics{~\cite{rubinov2009small}}.}

One way to identifying important design guidelines is through ranking studies. 
Cleveland and McGill  ~\cite{cleveland1985graphical} and  Mackinlay~\cite{mackinlay1986automating}
have characterized and ranked effectiveness by visual markers. In network visualizations,   
important nodes can also be quantified.
%, yet it remains unclear how to visualization quantitative attributes.  
For example,  %centrality indices 
 \textit{betweenness centrality} measures the  number of shortest paths between 
any  two  nodes that  pass  through this node~\cite{brandes2001faster}. As a result, 
betweenness centrality identifies \textit{hubs} (nodes with  high degree), 
\textit{modularity} (a measure of the  organization in modules  with  high  clustering), 
and  \textit{hierarchy} (a measure of how hubs  are connected in space)~\cite{wang2010graph}.  
Other  centrality indices  are also  useful.  
For example, the  \textit{degree centrality} of a node  is proportional to  the  \textit{degree} or the  
number  of edges  linked  directly to it and thus  reveals  the neighboring relationships. 

Despite recent great  strides in computational solutions, design guidelines for aggregating quantitative data 
on  networks are  few~\cite{alper2013weighted}. Ghoniem, Fekete, and Castagliola reported 
that the accuracy (i.e., the percentage of correct answers) was 
around $25\%$-$55\%$ using force-directed layout  and increased to $55\%$-$75\%$ with  
matrix  for tasks 
%of finding the 
%degree centrality
of 50-100-node  networks{~\cite{ghoniem2004comparison}}. This  work did  not  treat  node  connectivities 
as  a visual  variable and participants must visually compare the quantities through graph-node-positioning 
techniques. The study reported here instead shows empirically that  direct encoding of quantitative measurement  
can increase accuracy to $\geq 80\%$ for networks with  similar complexities. 

The  graph drawing community has  demonstrated  extensively that  node  and  edge  esthetics affect efficiency and 
effectiveness. Positioning techniques influence graph reading effectiveness~\cite{lam2007overview}: one should reduce 
overlapping edges and nodes and  avoid long  and  convoluted
edges~\cite{purchase1997aesthetic, vehlow2015state, huang2016evaluating}. Recent studies also address important issues  of temporal data  visualization~\cite{beck2016taxonomy} and  preserving mental maps~\cite{archambault2011animation}. 
A survey of specific  network uses  in brain  imaging of functional magnetic resonance imaging (fMRI) network visualizations reveals  that  coloring  is the most  common encoding solution~\cite{christen2013colorful} even though 
it may not  be optimal for  showing quantities~\cite{card1997structure, maceachren2012visual}. 
One  reason is that  there are few  guidelines for 
overlaying markers over networks to encode attributes.

The goal of this work  is to answer the following questions: 
\textit{How  much improvement can a direct  encoding of node  
attributes of quantitative measures make?}
\textit{How various positioning or layout are ranked?}  
\textit{Which encoding improves the performance the most and what is their ranking?} 
This study focuses  on the effects of network layout (or positioning; three  levels)  and  visual markers (nine  types)  
in representing quantitative data  on  the efficiency and  effectiveness of multi-scale comparison
tasks (Table~\ref{tab:taxonomy}). We have made use of the aggregation technique called 
\textit{attached-juxtaposition}~\cite{vehlow2015state},  where  attributes  are  displayed side-by-side with the graph. 
We 
%compare nine visual  markers  and  three  positioning techniques  and  
examine marker efficiency for tasks related to three aspects:  
\textit{ease of seeing network comparisons at scale},  
\textit{ease of network traversal}, and  
\textit{multi-scale marker distinguishability}. 
We have used  brain networks to guide our tasks to associate data and task scales to 
real-world uses.

%We have compared the usefulness
%of three positioning and nine visual marker choices to understand the strengths 
%and weaknesses of overlaying quantities over networks.

Given our clinical setting, we chose a gray background with white graph features, as shown in Table~\ref{tab:taxonomy}, 
which is suitable to dim medical exam rooms. It could be argued that black features on a white background would be a more 
common and generic setting, but our pilot study with six participants revealed no statistically significant differences between 
these two color schemes (Appendix A includes the pilot study results.)

%While the gray-background 
%shown in   used in this study
%is suitable to a dim medical exam room in clinical settings and one would argue 
%white background would perhaps be the most common solution.
%Our pilot study with six participants revealed no significant differences between white and
%gray in the study and
%markers were visually distinguishable in all trial conditions.

Our contributions here include:

\begin{itemize}
\item
A network visualization taxonomy including layout and  visual  marker selections;
\item 
An  empirical evaluation of positioning and  marker effectiveness in  network comparison tasks;
\item
A characterization of visual  markers that may explain  the performance differences; and
\item 
Design  recommendations for aggregating quantities in  network visualizations based  on  network structures  and  marker effectiveness.
\end{itemize}

\section{Background and Related Work}

This section  discusses two challenges that have influenced the choices of 
connectivity visualization methods: positioning and  network   properties  visualizations.
We  follow  Henry,  Fekete, and  McGuffin~\cite{henry2007nodetrix} in using  ``graph'' to refer 
to the topological structure without  associated attributes and  ``network'' for  a  graph with  
an  arbitrary number of  attributes  associated   with   its  vertices   and   edges.   
For  matrices, ``rows'' and  ``columns'' are the visual  representation of vertices  and ``cells'' 
are the visual  representation of edges.

%\begin{comment}
\remove{
{\subsection{Background in Brain Functional Connectivity}}
The brain structure contains networks over a hierarchical structure, 
where the left and 
the right hemispheres are at the highest level followed by six lobes: frontal,  parietal, occipital,  
temporal, central,  and  limbic
lobes{~\cite{strominger2012gross}.} The  next  level  is  cortex  
%on  the  sulcus and  gyrus 
where brain  scientists search for differences in brain connectivities
by extracting structural correlations{~\cite{irimia2012circular}}. 
%the correlations are computed in fMRI studies. 
Our study makes use
a 74-node network of the cortex and sub-cortical brain regions in the Network Based Statistic Toolbox sample 
data{~\cite{zalesky2010network}}. }
%\end{comment}

\subsection{Node Positioning}

The  first  challenge in  network visualization is  to  choose the  node  positioning or  layout algorithms, often  with  
the goals  of  placing nodes to  represent groups, child-parent relationships, adjacencies, or relative physical node  
proximity, and other aesthetic criteria.   
Nodes belonging to  the  same  group or adjacent neighbors  are  placed closer.  
Irimia et  al.~\cite{irimia2012circular}  use Circos~\cite{krzywinski2009circos} for  circular   layouts that  facilitate   
the  display of relationships between positions by links and heat maps. 
Concentric circles  are  used  to  present the subdivision of the  hierarchical structures that is also  adopted in our  study. 
The matrix  views  show  connectivities  on  a  grid   on which  connections are  labeled in cells; these matrix  
representations can reveal  both local and global  connectivities~\cite{marcus2013human, mori2013atlas} and  
are used  in comparison studies to differentiate two node  attributes~\cite{alper2013weighted}.

\remove{
three-dimensional (3D) positioning 
techniques  associate  nodes  with anatomical 
landmarks{~\cite{gerhard2011connectome,  bassett2016small}} and are thus essential for 
accurate illustration of relative anatomical proximity of network nodes. However, 3D 
introduces visual clutter in showing dense \remove{brain} connectivity{~\cite{irimia2012circular, chen2012effects}}; 
Sophisticated interaction techniques are required to 
filter the data and orient the users, e.g., through edge bundling{~\cite{bottger2014three}}
and interaction in 3D.
}
\remove{Two-dimensional (2D) projection-based visualizations{~\cite{bassett2011conserved, xia2013brainnet}}  
are more popular compared to 3D techniques. 
These approaches do not concern 3D locations and follow the 
low-dimensional representation of graph theory to show structure \change{of different brain regions}{in data}.} 
%However, using many  such visualizations  creates dense node clusters throughout the 3D spatial extent  of the connectivity representation and makes it difficult to discern the underlying structures.
%  in a top-down approach (e.g., brain hemisphere to cortex  to lobes  
%to gyral / sulcal  structures).
\remove{Our study compares the circular, matrix, and also direct projection layouts as 
shown using a brain network data.}

%\add{ }

\subsection{Network Attributes Visualization}
The  second challenge in  network visualization is  to  represent   vertices   or  edge  properties. 
Often  the  goal  of  a visualization is to maximize the  information content while minimizing superfluous graphical elements, 
i.e.  to  preserve  a  high  data-to-ink ratio~\cite{tufte1983visual}. Among many marker encoding designs 
for quantitative data,  color, node  size, and edge  width may be the most  popular approaches. For example, the connectome viewer  toolkit~\cite{gerhard2011connectome} and Bassett et  al.~\cite{bassett2011conserved, bassett2016small} represent 
connection strengths with  edge thickness to demonstrate the  small-world attributes. 
However,  edge  thickness may not  be scalable,  especially when edge  density becomes  so great as to cause occlusion.

A recent  survey on  analyzing brain  network visualization reports that  $85.8\%-90.6\%$ of the  
images  in  published results use color~\cite{christen2013colorful}. However, understanding of coloring in visualization remains limited. We know that coloring generally facilitates visual  grouping~\cite{laplante2014connectome, alper2013weighted}, although common sense  suggested  that  coloring  could  be poor  for show- ing  quantities.  However,  many   coloring   design depend on  application area,  and  our  current study also  showed that  the many  hues  in the  color  
map  of Kindlmann, Reinhard, and  Creem~\cite{kindlmann2002face} assist  comparison and  achieve the  most accurate results.
% for comparison tasks.
%In our study, we expand the study of vertices attributes using nine marker types. 
%Node size  is  also  used  to show degree information in  3D  visualization.
%Color maps  are applied to the several  circular  dimensions to show curvature, volume, gray  matter thickness, centrality and 
%other  measures available in the  LONI  pipeline environment~\cite{rex2003loni}. 
%Besides color maps,  

% for showing quantitative information into this empirical study. To study marker effectiveness, we extend the  encoding choices from color and  size to include texture, shape, orientation, and slope. 

\subsection{Network Effectiveness}

\remove{The design of network encodings has  focused in the past decades on aesthetic attributes, leading to 
great  insights on  criteria  
to  improve readability{~\cite{purchase1997aesthetic, purchase2002empirical}}. }

Node   positioning  affects  human accuracy   in  judging node connectivities~\cite{ghoniem2004comparison} and  
quantitative marker effectiveness for set or grouping tasks~\cite{saket2014node, jianu2014display}. Overlaying colors  
on nodes and   explicitly showing  groups  by   drawing  boundaries have  been  found necessary, as  have  other  properties such as  \textit{symmetry}, \textit{spatial proximity}, and 
\textit{associated set} relationships~\cite{alsallakh2016state}. 
GMap~\cite{gansner2010visualizing}, BubbleSets~\cite{collins2009bubble} and  
LineSets~\cite{alper2011design} address  domain-specific grouping. Compared to these  
important grouping studies, our study includes a set of new comparison tasks  not  directly considered in  previous work. 
First, here we study the overlay of an extensive set of visual markers on  nodes. 
Second,  quantitative data   comparisons offer design guidelines for comparisons at different spatial distributions, 
such as within closely proximate regions (e.g., nodes belonging to a set), 
while  others  may  be distributed in  space  (e.g.,  neighboring nodes).  
The  proximity differences perhaps influence context and thus layout effectiveness 
when the  nodes to  be  compared in  close  proximity would lead to better  comparison performance and further away
nodes count on more effective visual marker solutions. 
As a result, markers supporting effective comparison must  be distinguished for designers to make  sound design choices.

%\remove{, which  
%is associated with  node  positioning techniques.} 
%{Recent studies also  extensively address  grouping}
%Colors have  been  studied extensively, but, other  visual markers (e.g., size)~\cite{mackinlay1986automating, maceachren2012visual} may be more  suitable to quantitative data encoding. It is therefore necessary to study the positioning and the quantitative data  attributes together, as we do here.

In the area  of quantitative attribute comparisons in complex real-world  scenarios, the  empirical studies most similar   
to  ours are performed on  geospatial data  where nodes are  also  spatially distributed (e.g.~\cite{garlandini2009evaluating}). 
For  example,  Garlandini and  Fabrikant compared size,  color  value (lightness), color hue,  and  orientation for change 
detection, whether or  not  a change exists and if so where and  of what  kind, and  discovered rank  order differences. 
That  study found that,  consistent with the  2D marker ranking of Mackinlay~\cite{mackinlay1986automating}, size was most and  
orientation was least accurate marker for quantitative data  and  hue  was  most  visually salient{~\cite{garlandini2009evaluating}}. 
Their  spatial context  is largely, however, not a network but a map  and the  tasks are also different{~\cite{fabrikant2004distance}}. 
The applicabilities of their  design guidelines to networks need  to be reassessed because graph efficiency is affected by the 
presence and absence of edges{~\cite{saket2014node}} as well as the size and  
density scales{~\cite{ghoniem2004comparison}}, symmetry and  orthogonality{~\cite{purchase2002metrics}}, and  
rendering{~\cite{tory2009comparing}}.

Two domains of network use have to our  knowledge benefited from important quantitative markers in network 
uses:  genomics~\cite{gehlenborg2010visualization} and  connectomics~\cite{lichtman2014big}. 
Understanding multi-attribute visualization in genomics and brain  networks has  recently been  proposed as a leading design challenge. Here we follow  up  on  pioneering work  in fundamental graph and  geospatial studies
% to determine  the  
%relative importance of different contexts  in some new  comparison tasks  and  data.  We  also  hope 
to  place the value  of overlaying quantitative data  over  networks on a firm empirical foundation by developing and  applying metrics  for measuring effectiveness and  efficiency.

\begin{comment}
Inspired by pros and cons of these solutions, we have studied tradeoffs of
marker designs and situated these design in the brain connectivity analysis
to study whether or not we could achieve scalable solutions of encoding network attributes on
nodes. 
Our current study explores the influence of 
\textit{positioning} (projection, matrix,  and  circular  layout) and \textit{visual markers} on efficiency, effectiveness, and 
subjective preferences. 
\end{comment}

%Similarly   to  their   work, we  also  rank   attributes and   focus  on  whether or  not  a particular design fits  its  
%real-world uses.  We  must  realize that  their  tasks  are  naive  visual   search   driven by  visual attention where the 
%participants had   no  a  priori   knowledge  of the location of the  target,  while  ours  were  primed: participants 
%knew  the  target  locations through interaction.  In  addition, principles in  geoscience for  studying distributed 
%quantitative attributes may  not  be  applicable to  brain  networks with  edges,  attributes, and  hierarchies. 
%Saket  et  al.  have  empirically demonstrated  that  task  performance differed significantly after  adding links and  
%areas to a set of scattered nodes{~\cite{saket2014node}}. Finally, the change-blind test uses different working memory
%mechanisms as our visual search and ours is largely influenced by visual memory contexts{~\cite{brady2011review}}.

\section{Experiment}

\begin{figure}[!t]
	\centering
	\includegraphics[width=\linewidth]{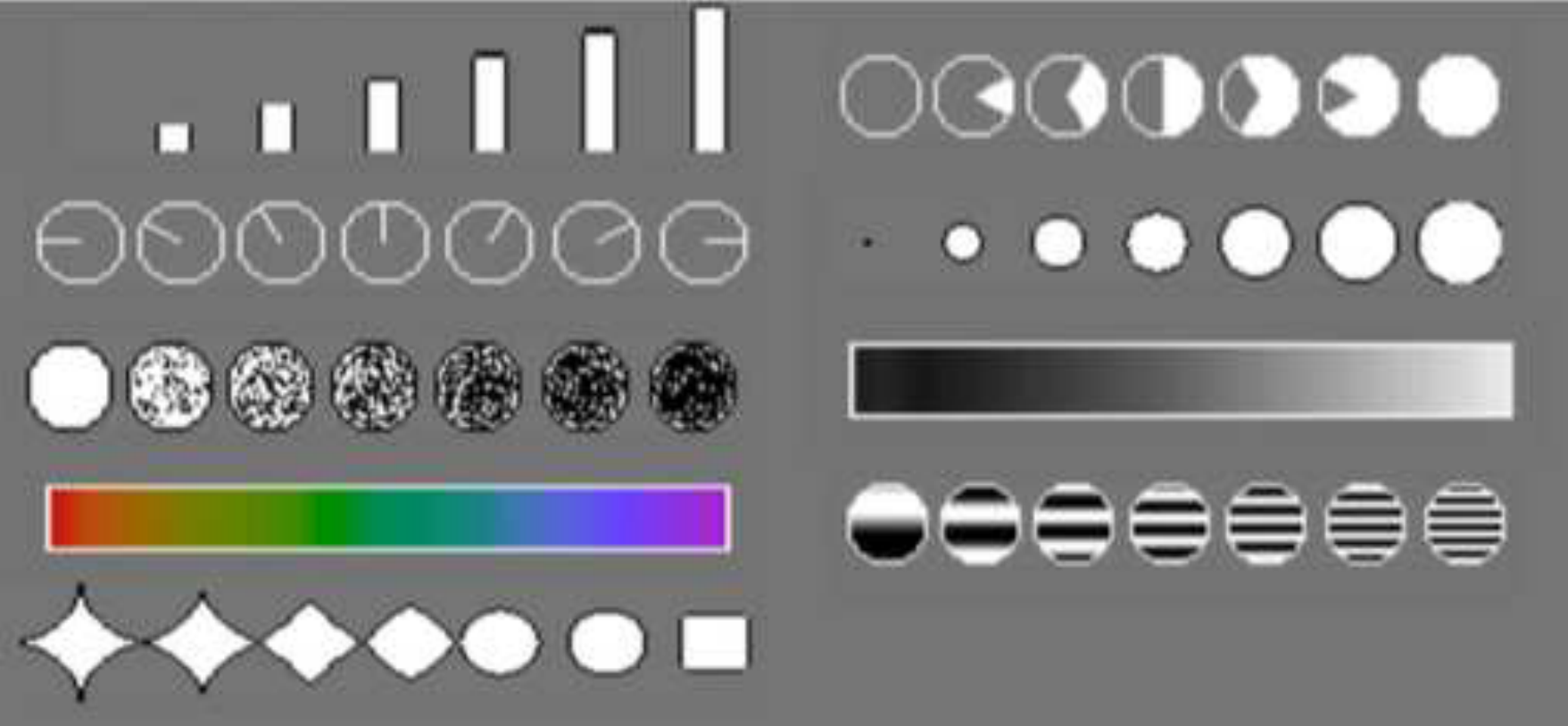}
	\caption{Nine Marker Types.}
	\label{fig:taxLegend}
\end{figure}

Our  experiment used a within-participant design with  two independent variables: \textit{layout} (three  levels: projection, circular, 
and  matrix) and \textit{visual markers} (nine levels:  length, shape, density, slope,  angle,  size,  texture, hue,  and  
lightness) (Fig.~\ref{fig:taxLegend}). 
Dependent variables include accuracy, task  completion time,  and  subjective ratings. 
%four multi-scale comparison task types.  Visual markers encode the centrality measures.

\subsection{Domain-Specific Data Selection} 

Tasks and datasets derived from real-world data provide the best and most realistic foundation 
for relevant and generalizable findings~\cite{wolfe2016use}.
%~\cite{chen2009domain}. 
For this reason, we draw on our existing expertise and collaborators to base our study on data and tasks 
from brain connectivity research. However, these tasks and data are still generally valid across many 
applications relevant to networks.

%We  have  used   brain   connectivity datasets  to  inspire our study because tasks and data  abstracted from real-world data 
%but reflecting real-world complexity and  design principles give  fundamental guidance on network uses,  inspired by 
%use-based basic  research and  domain characteristics  that  could  be  useful  might otherwise be  
%missed in  generic
%studies. The  brain  structure contains networks over  a hierarchical structure, that permits us to 
%understand a set of complex networks of relationships at scale.

Our  study uses 27 brain  network cohorts  (15 normal and  12 diseased brains)  in the 
Network Based Statistic Toolbox sample data{~\cite{zalesky2010network}}. 
The left and  right  brain hemispheres contain a hierarchical structure: 
%six lobes
%each:  frontal, parietal, occipital, temporal, central,  and limbic lobes{~\cite{strominger2012gross}}; 
four lobes (frontal, parietal, temporal, and  occipital  lobes)  are used in this dataset and our study. 
The  next level  is  the 74 cortex  and  subcortical brain regions where differences in brain connectivities are extracted  
by structural correlations{~\cite{irimia2012circular}}. 
%Each  sample  is measured using  
%74 brain regions in the four brain lobes. 
Edges (connectivities) between nodes are  defined by  all  pair  correlation strengths of 
$74\times74$ numeric values.  These  correlations can be either positive or negative. Here  we use  correlation strengths, i.e. the absolute values of the correlations.

%We have also positioned the nodes considering domain-specific attributes such as symmetry and brain
%hierarchical structures and measured meaningful tasks at a scale practical to real-world applications. 

\subsection{Independent Variables: Positioning and Visual Markers}

We describe a taxonomy to design quantitative networks with three positioning and nine visual markers (Table~\ref{tab:taxonomy}).

\subsubsection{Three Positioning Techniques}

\textbf{Projection.}  
%The superior projection shows  the front  of the brain from above. 
The project shows a 2D projected view. 
A node  layout algorithm overlays a placement grid  
and places  nodes and  labels  at the  nearest available grid  point, as  in  Chen  et  al.~\cite{chen2004testbed}. 
A light-gray bounding box contains all nodes belonging to the same parent. 
Some parents  (e.g., parietal and  temporal lobes) overlap 
due to the  2D projection, but  can be differentiated by the bounding box and by hovering the mouse pointer over 
the label 
%the  lobe  label  
to  highlight all sibling nodes.
% within that  lobe.  
We apply geometry-based edge  bundling~\cite{cui2008geometry} to 
reduce visual  clutter caused  by overlapping edges.

\textbf{Circular.} 
The circular  layout reflects  the  symmetry and hierarchy in data.
%of the brain anatomical structure.
Following the design in Irimia   et  al.~\cite{irimia2012circular},
the  left  and   right  symmetrical structures 
%hemispheres 
are symmetrically placed on the  left and  right  half-circles.  
The four parent nodes are ordered in 2D from  top  to bottom following their anatomical locations in 3D.
% as frontal, parietal, temporal, and occipital  lobes. 
%The text labels in the circular  layout are parallel to the radial  direction. 
Hierarchical edge  bundling~\cite{holten2006hierarchical} of the connectivity shortens edges within the same lobe compared to those  between different lobes.

%by curving the links between sub-hierarchies to 
%their least common ancestor in the hierarchy. The hierarchy is defined by lobe clusters. Each lobe has a invisible dummy node representing the root that  is placed at the center  of the angular position with  half the radius. Thus links connecting nodes  representing regions in the same  lobe are shorter and  less curved, while  inter-lobe links  are  more  curved to the  center  of the  circle and  are longer.  In this  way,  regions within and  between lobes can be clearly  differentiated.

\textbf{Matrix.}  
The  order of nodes in  the  matrix  is  the  same as  those   in  the  circular   layout in  order to  preserve the symmetric 
structure and  the order in the anatomical regions. Edge  connectivity between two  nodes is shown in  solid white  in the matrix  cell. 
The small  cell sizes challenge the labeling, however, as noted by Alper  et al.~\cite{alper2013weighted}. We label and 
display the attributes by placing them  beside  the  rows  and  columns in a jigsaw pattern; i.e., the  markers except  the  length  
are  arranged in two  lines  so that  each  marker can occupy  two  cell widths. The length  bars are not constrained by space  
issues  and use  a single  column, as in Henry et al.~\cite{henry2006matrixexplorer}. Labels  are placed next  to node  markers and
are displayed vertically at the top of the view and horizontally on the right. In cases where 
two  matrices are displayed side-by-side, the matrix  on the right does not display any  labels  since  they  can  be  shared with
the  labels  on  the  left.  White  horizontal and  vertical  lines separate the lobes.

\subsubsection{Nine-Centrality-Attribute Encoding}
We  select  nine  visual   markers to  encode node   attributes (here  centralities): length, angle,  slope,  area,  density, light- ness, hue,  texture, and  shape.

\textbf{General criteria for marker size}.
We   follow   the
psychophysics literature  on  marker distinguishability in choosing  marker sizes and  use the same  size whenever possi- ble to ensure fair comparison of these  nine  marker types. Human comprehension of  high  spatial frequencies can reach  eight  to  ten  degrees per  cycle  without causing  visual   blurring{~\cite{barten1999contrast}}.  
Here   we  select  the  texture marker to  contains at  most  ten  cycles.  This  indicates that marker size could  be as small 
as $1^\circ$. When  a viewer  is about 400mm  from  the  screen,  the  marker size  should be 7mm, so
we fix circular  markers at 7mm diameter. 
%We use 50\% gray  as the background color and  outline the markers in black to increase contrast and a pilot study
%showed that there was not significant differences between the background colors.

% (except  when  the size is explicitly 
%defined in the length and  area encoding.)
%This gray background is used  in all visualizations.

\textbf{Length} and  \textbf{area}
are  size  encodings. Length maps  the value  linearly to the  height  of a bar.  The maximum length is mapped to 20mm  and  the width of the bar to 2.5mm, given  the screen  space.
%The bar width is $0.4cm$, equal  to the radius of the  circular marker. 

\textbf{Angle} and \textbf{slope} are  orientation encodings. The  angle encoding resembles a pie chart  in which  the angle  
indicated in white  is proportional to the underlying numerical value. We  render the  area  within that  angle also in  white.  
The  slope encoding  uses   half  circles  ranging from $0^\circ$ to $180^\circ$.  
This clock-like  marker starts  by pointing to the  left and  rotates clockwise  to its maximum pointing to the right.

\textbf{Density} 
is defined as the amount of black on white.  One-pixel-wide black  dots  are  randomly placed in the  circular mark and  the number of dots  is linearly mapped to the centrality indices.  For the experiment monitor of a $23''$ $1920\times1200$ resolution display with  pixel size $0.258mm$,  a mark  of  $7mm$ diameter has  area  $38.465mm^2$, which  contains approximately 578 pixels.

% with a maximum value of 500 possible locations. To avoid  overlapping dots,  the 500 possible locations of the  
%dots are  precomputed and  evenly distributed in the circle. Then N locations are 
%randomly selected from these 500 locations. 
%So a marker with max density should have approximately 86.5\% of the pixels rendered in black and the rest in white.
% based  on the encoded value  and  black dots  are drawn at these locations. 

\textbf{Lightness} and \textbf{hue}
are two color components. Lightness goes from  black  to white  and  is linear  to the  centralities. The hue  markers use 
the isoluminant map  in Kindlmann, Rein- hard,  and  Creem~\cite{kindlmann2002face}. 
Though hue,  unlike  lightness, is not perceived as ordered~\cite{chung2016ordered}, 
using  a large spectrum could  help distinguish small  differences in encoded values  (depending on the color resolution) and  
facilitate  fast comparison.

%\change[gz]{using a \textit{log} function following Stevens's power law{~\cite{stevens1966brightness}}}{By mapping the value linearly to the lightness component in the LAB color space. We pre-computed 20 colors for 20 evenly distributed points in the LAB color space where $a=b=0$ (i.e. gray scale) and lightness change from 0 to 100. We then interpolote the values between precomputed points to get the final color (Dr. Chen, if I understand it correctly, Stevens's power law does not apply here because it works with physical stimuli, i.e. how much luminous intensity (unit: lux) does the pixels on the screen emit, which should be measured using. This is not the same lightness concept as that in the color space. )}. 
%\add[gz]{It provides six control point colors and we interpolate between them to generate a continous mapping.} 

\textbf{Texture}
is  defined  by   frequency  of  black-white
strips   to  indicate encoded centrality, following methods in Bertin~\cite{bertin1983semiology} 
and MacEachren~\cite{maceachren2012visual}. 
Frequency increases linearly with   centralities to  a  maximum of $10-cycles/degree$.

\textbf{Shape} 
is  defined using   the  ordered glyphs of  2D  superellipses~\cite{schultz2010superquadric, seltzer2016glyphs} generated 
by the following parametric equations: 
$x(\theta)=A|\cos\theta|^\alpha\cdot sgn(\cos\theta)$, 
$y(\theta)=A|\sin\theta|^\alpha\cdot sgn(\sin\theta)$, where $sgn(t)$ = -1 ($t<0$) or 0 ($t=0$) or 1 ($t>0$), 
A is the radius, and $\theta$  is the angular position of a vertex  on the marker's boundary, ranging 
from  $0^\circ$ to $360^\circ$. 
Hence  the  shape  morphs gradually from  a  near-star to  a near-square where the  curvatures of  the  
shape   boundary vary to show  ordering in the data~\cite{forsell2005simple}. 
We use  a \textit{log} function to map  the centrality data $v$ to  the  control   
parameter $\alpha$, where  $\alpha=a\log(bv+c)$ and $a=-6.64$, $b=0.55$,  and  $c=0.3$ 
are empirically chosen,
so that  the entire  centrality range  is mapped to a visually differentiable curvature range.  For each $v$, 
the marker area is normalized to the circular  area in other  visual  markers by changing the marker's diameter $A$ 
to avoid   confounding area cue.

\subsection{Tasks}

\begin{figure}[t!]
	\centering
%	\begin{subfigure}{0.95\columnwidth}
		\centering
		\includegraphics[width=\columnwidth]{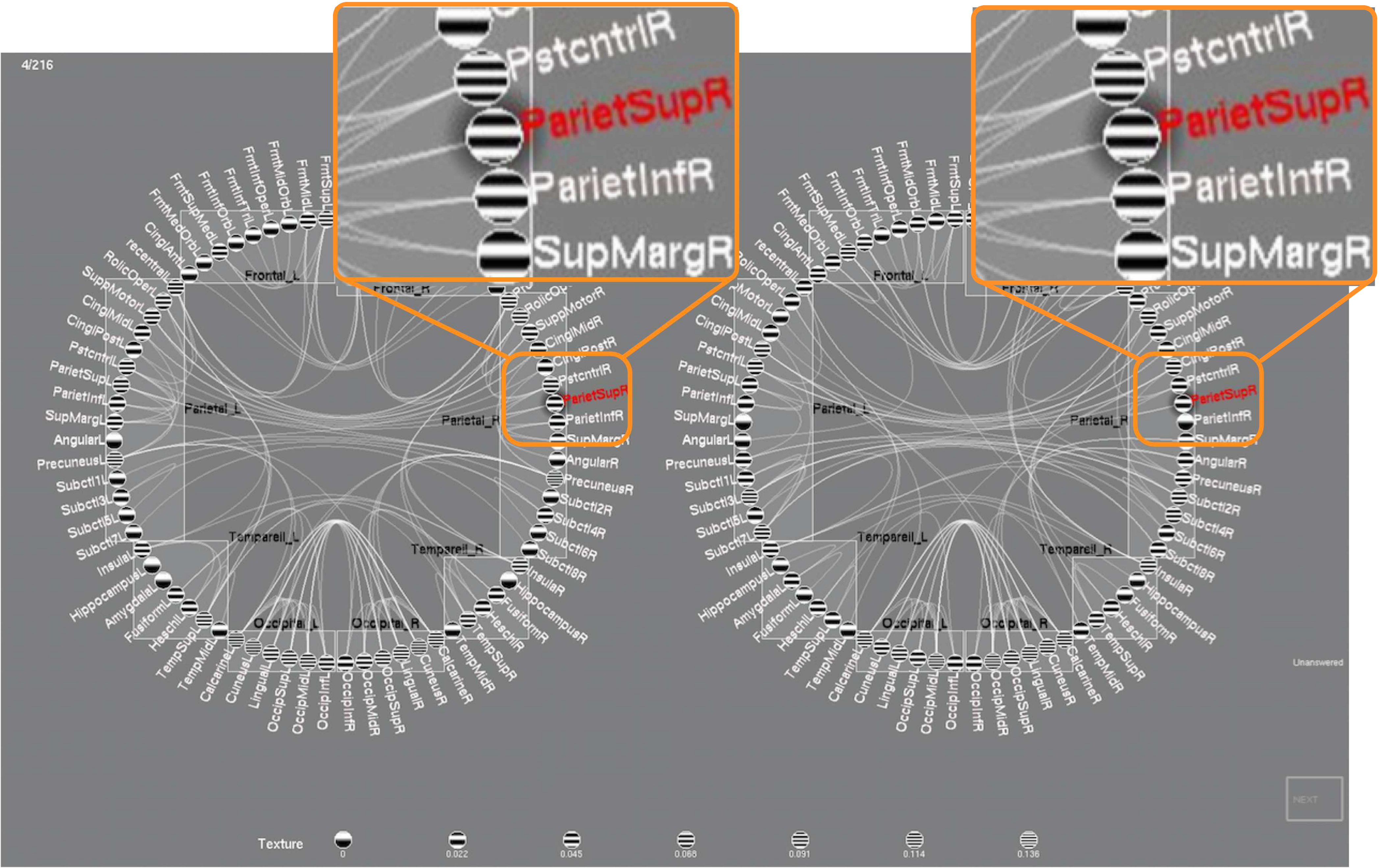}
		\caption{Task type 1 (Change-detection): \textit{In which graph does the highlighted node have 
		a higher degree of centrality?} This example
		uses circular-texture.
		% and the key is the node in the left figure.
		}
		\label{fig:tasks:1}
	%\end{subfigure}%
\end{figure}

\begin{figure}[t!]
	%\begin{subfigure}{0.95\columnwidth}
		\centering
		\includegraphics[ width=\columnwidth]{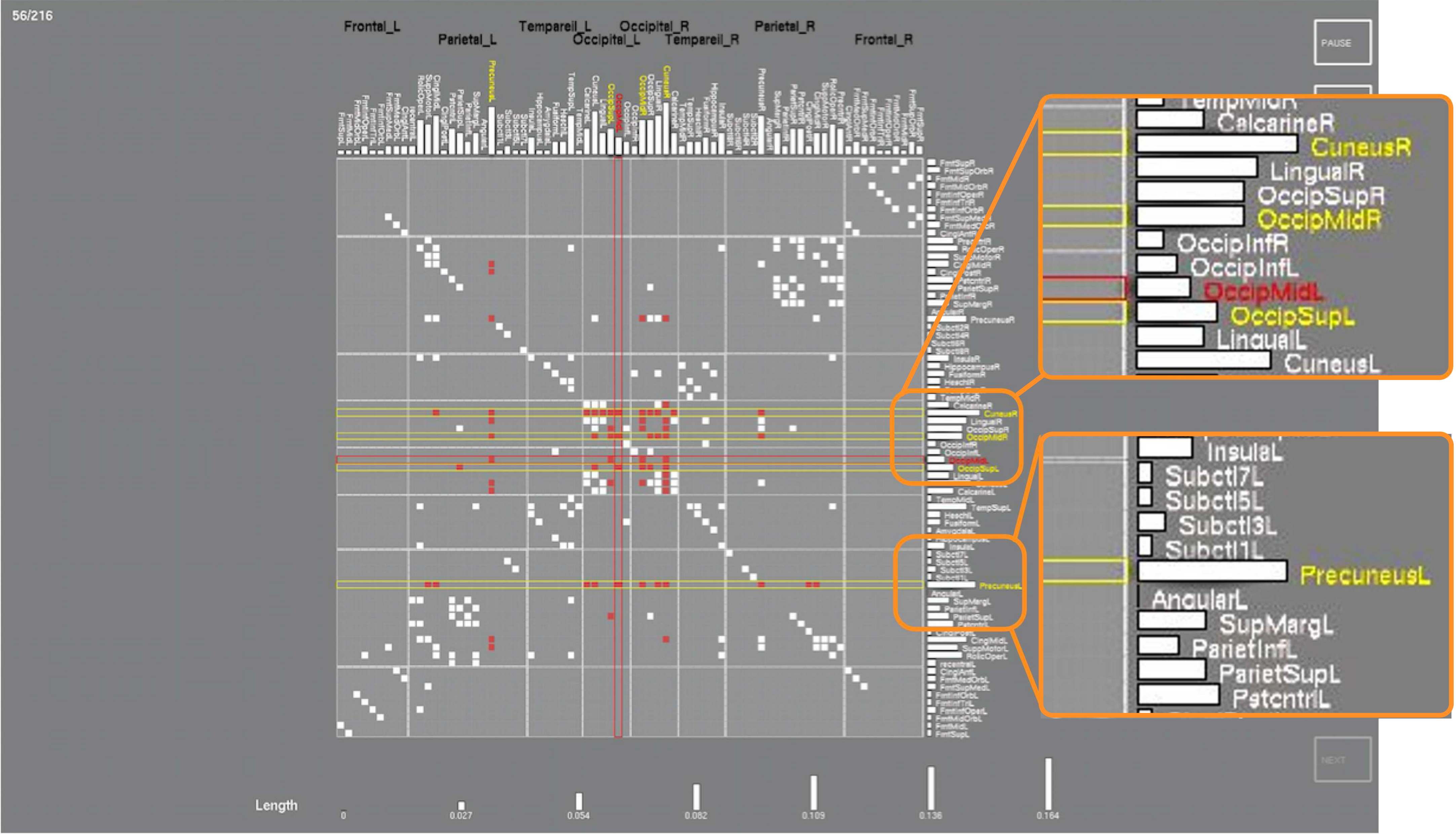}
		\caption{Task  type 2 (NeighborHub): \textit{Among the neighbors of  the highlighted node, which has the highest betweenness 
		centrality?} This task example uses matrix-length encoding.
Among the four neighbors (CuneusR, OccipMidR, OccipSupL and PrecuneusL) of OccipMidL, CuneusR has
the highest centrality. }
		\label{fig:tasks:2}
%	\end{subfigure}%
\end{figure}

%This participant first highlighted all neighboring nodes shown in yellow and then compared these
%values. 

\begin{figure}[t!]
	%\begin{subfigure}{0.95\columnwidth}
		\centering
		\includegraphics[ width=\columnwidth]{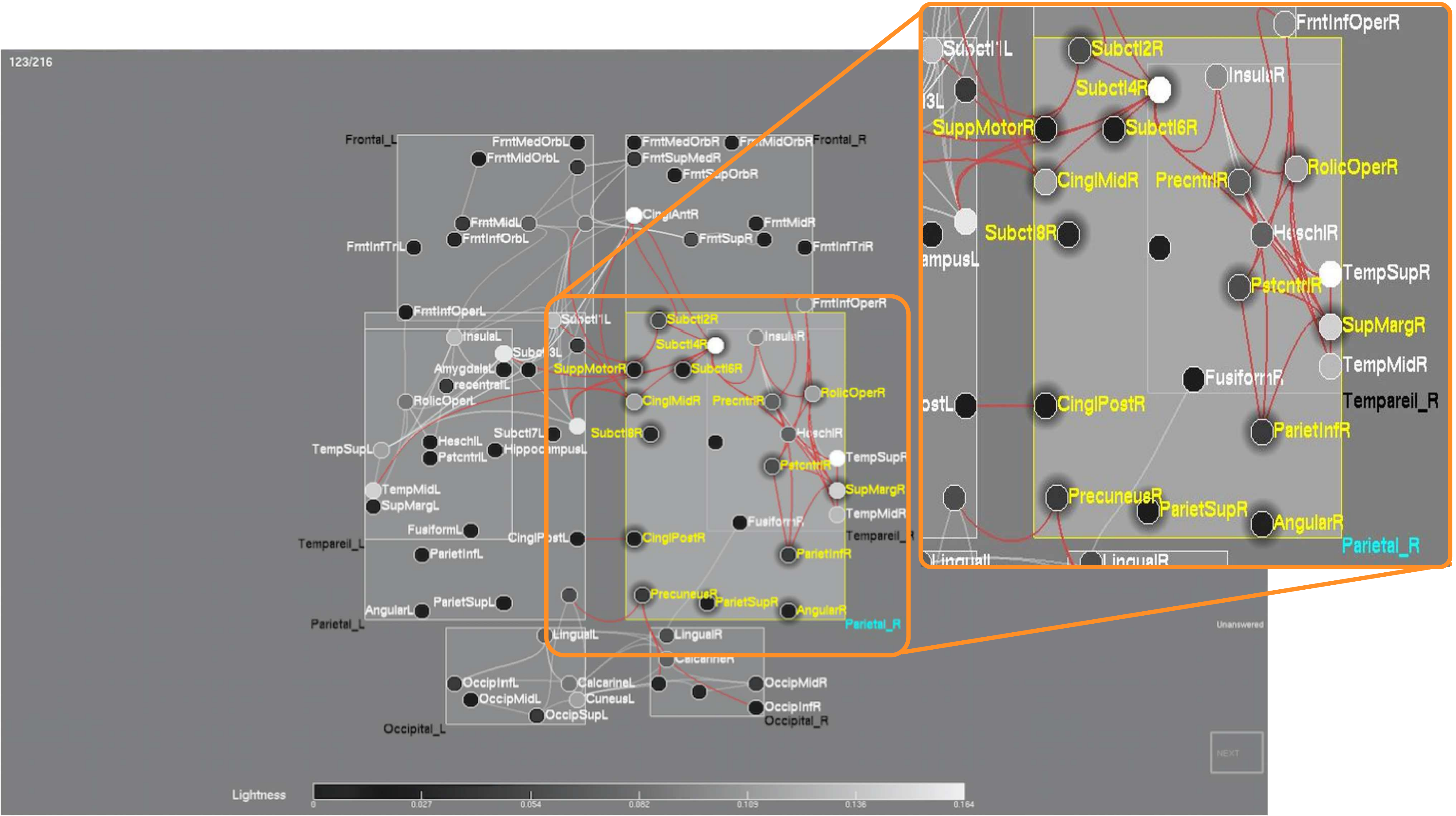}
		\caption{Task  type 3 (LobeHub): \textit{Of all the nodes in the highlighted brain lobe, which has the highest centrality?}
		This example uses projection-lightness.% and the correct answer is Subctl4R.
		}
		\label{fig:tasks:3}
	%\end{subfigure}%
\end{figure}

		%The participant first highlighted all nodes in the ParietalR and then compared the centrality of all nodes in that lobe. 
		
\begin{figure}[t!]
%	\begin{subfigure}{0.95\columnwidth}
		\centering
		\includegraphics[ width=\columnwidth]{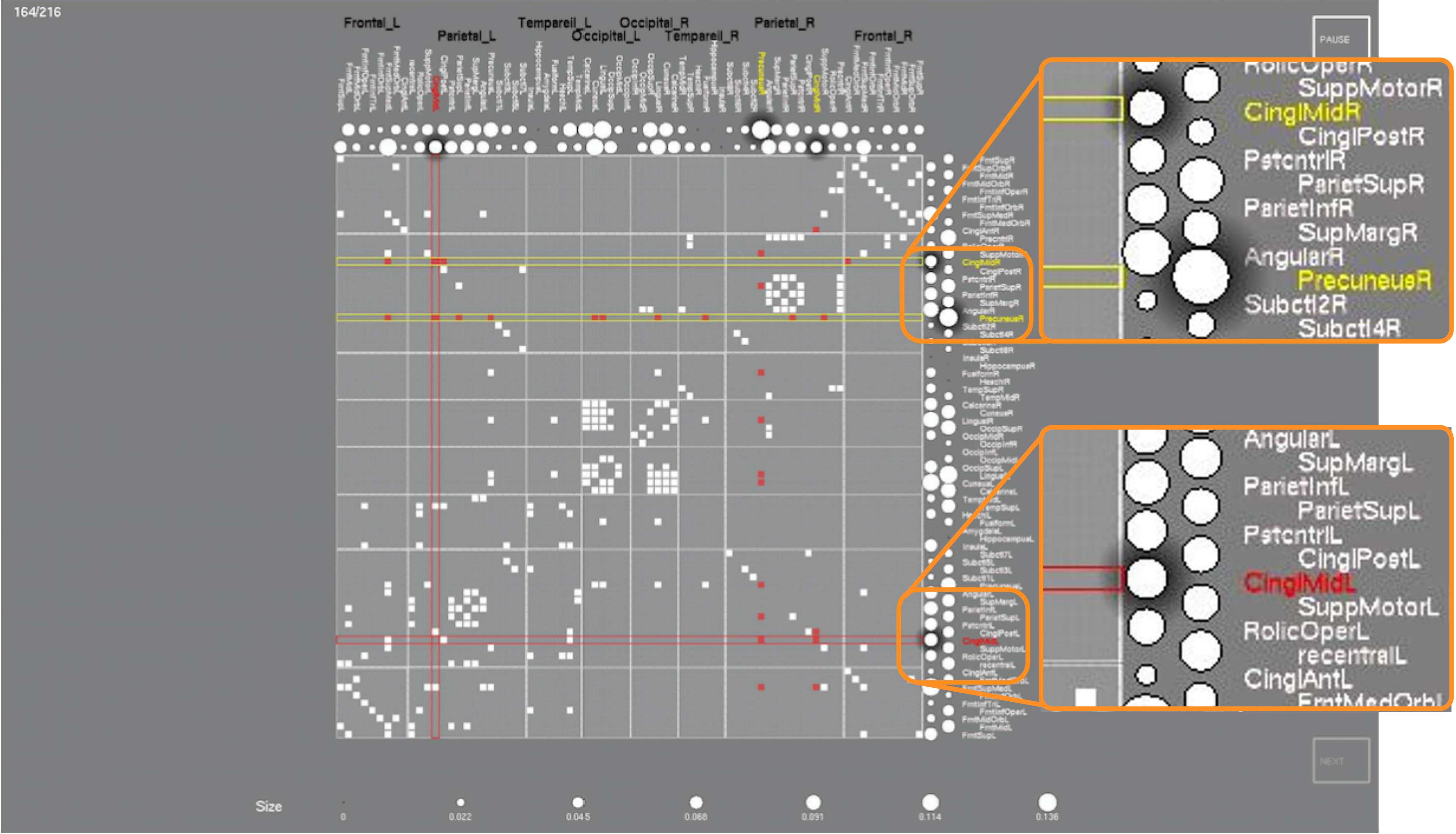}
		\caption{\small Task  type 4 (HemisphereHub): \textit{Of the neighbors of the highlighted node, 
		which one has the highest centrality in the opposite hemisphere?} 
		This example uses matrix-area. Among the neighbors of the node CingMidL in the opposite 
		hemisphere (CingMidR and PrecuneusR).	% the correct answer is PrecuneusR.
		}
		\label{fig:tasks:4}
%	\end{subfigure}%
%	\caption{Four Tasks in Our Empirical Study.}
\end{figure}

%	\label{fig:tasks}
%\end{figure*}

Four task types are selected  to measure the effectiveness and efficiency  of positioning and  encoding techniques. The first task  is a change-detection task  while  the  last  three  tasks address hub  finding in local and  global  networks involving various scales
of network tracing and  comparison.

\textbf{Task  1 (Change-detection)} (Fig.~\ref{fig:tasks:1}). 
\textit{In which graph does the highlighted node have a higher degree of centrality?}  
This task  is a change-detection task  asking  participants to compare centrality changes directly in  two  networks. The two  network visualizations are placed side by side. To select  the  answer, participants use  the  mouse to right-click
one  of the  two  networks to  indicate the  one  with  higher centrality.

%Here the proximity of the information and  the marker discriminability determine task performance.

\textbf{Task  2 (NeighborHub)}  (Fig.~\ref{fig:tasks:2}). 
\textit{Among the neighbors of  the highlighted node, which has the highest betweenness centrality?}
This task asks participants to find the hub within a set of nodes that  are connected to the node  highlighted in red.  
To complete this task, participants must  first find the neighbors of the  task-node through edge  tracing, then  compare the 
centrality measures of all neighbors to locate  the  largest  one,  and then  right-mouse click to any part  of the 
answer node  (e.g., node or node  label) to  mark the answer.

%This  task is related to identifying the grouping function based  on  their  spatial closeness;  the  neighboring node  with  the highest betweenness centrality is the one that has the shortest path to the rest of the network.

\textbf{Task 3 (LobeHub)} (Fig.~\ref{fig:tasks:3}).  
\textit{Of all the nodes in the highlighted brain lobe, which has the highest centrality?}  
This  task requires participants to find the node  with  highest centrality  of all nodes within the lobe with name highlighted in  red.   To  complete  this  task,   participants first  need   to recognize the  hierarchical relationships, i.e., find  all nodes contained in a lobe,  and  then  compare the  values of those nodes. 
%Participants again  can select any  part  of the answer node  to mark  the answer. 
%In circular  and matrix  positioning, these  nodes are  side by side; in the  projection view,  a few
%lobes may overlap  due  to  2D
%projection and we have explicitly depict the bounding boxes of all lobes.

\textbf{Task 4 (HemisphereHub)}  (Fig.~\ref{fig:tasks:4}). 
\textit{Of the neighbors of the highlighted node, which one has the highest centrality
in the opposite hemisphere?} This  task  requires participants to  find  the  hub  on  the 
other symmetry part connected to the  region.
To complete the  task,  the  participant must  use  symmetry to find the neighbors 
on the other side before finding the one with the highest centrality. 
%Participants right-click the node to select an answer.

\subsection{Interaction} 

We have  designed and  implemented interaction techniques to make  perceiving structures and  quantitative data  easier in  different layout methods. Hovering over  the  parent  node label  highlights all sibling nodes;  left-clicking a  node highlights all neighboring nodes;  left-clicking  an edge highlights  the  nodes the  edge  links  to. These  text  labels  of the highlighted nodes are shown in yellow  and  the highlighted nodes are  haloed to differentiate them  from  the task  node in  red.  During the  experiment, the  task  node   is  always highlighted in red but changes to cyan if the participant selects the task node.  To support multistage interaction, we also  let participants click on  a highlighted yellow  node  to see all edges  linked  to this node shown in cyan.

In  matrix  view,  the  associated column and  row  of the
task-node are  also  in red  or cyan  to help  participants find the neighbors and  overcome in part  the difficulty of tracing edges.  
These  task-related red  nodes  are always visible during task execution.

\subsection{Data}

Our data  are carefully chosen  based  on results of three  pilot studies: we  balance   the  size  of  the  network so  that   the results will  be above  guessing rate,  yet  not  so simple  that all answers will be correct.

\textbf{Edge density.} 
We sample the edges  (originally $74 \times 74$)
based on network efficiency, which is the ratio of the number of  edges   to  the  number of  all  possible edges   in  the  
network~\cite{bullmore2012economy}. We keep  the  $5\%$ (or 135 edges) sparse and  top  $10\%$ (or 270 edges)  
dense  edges  of all edges  in the two edge  densities; these  densities are the same as used in Alper  et al.~\cite{alper2013weighted}. However, the  total  edge  count  in our  study is nearly  four  times  that  
in Alper  et al.~\cite{alper2013weighted} because the number of nodes is doubled.

\textbf{Data complexity} 
is chosen  at three  different levels from low  to high so as  to cover  a broad range  of real-world conditions.   For  task  1,  the  degree  difference in  each  of paired networks is 1 or  2 or  3 for  both  dense  and  sparse networks. Since  the  dense   cases  have  nodes with  higher degree, the same degree difference results in smaller marker difference and  thus  be more  difficult  to distinguish.

%\change[gz]{For task  1, comparison of two  values, we first calculate the mean correlation of  
%each  of  the  74  nodes   in  27  datasets (15 controls and  12 patients), then  
%calculate all-pairs degree centrality differences between individuals  and  the mean.  The histogram of centrality differences are plotted and three levels  are chosen  from  three  distributions: between 0 to 25 (high complexity or hc), 25 to 75 (medium complexity or mc), and  75 to 100 
%percentiles (low complexity or lc). The dense ($10\%$) and  sparse ($5\%$)  networks are calculated separately.}

Tasks  2,  3,  and   4  measure  betweenness centrality to
define   hub  nodes. To  determine the  hub  candidates, we calculate the  distribution of  betweenness centrality of  all nodes 
in the mean  graph, computed by averaging all 27 matrices for each of the pairwise correlations, to determine the 
betweenness centrality ($bc$) thresholds. The  thresholds are $mean(bc) + stdev(bc)$ for both sparse and  dense  conditions; 
a node  is a hub candidate when 
$bc$ $\geq$ $mean(bc) + stdev(bc)$ for that  node.  The data  are chosen  from  the set of hubs.  To
ensure legibility, we also select hub nodes whose centrality is at least $5\%$ higher than any other  nodes that  
are also neighbors of the task  node  (for tasks  2 and  4) or are in the same  lobe  (for task  3), so that  the  markers 
encoding these values can be distinguished visually.

For task 2 on the NeighborHub tasks, since a pilot study shows  that  the  number of neighbors of the  task node strongly impacts task  performance, we  first  identify all hubs  for each network, then  examine all the neighbors of all these  hub  nodes and  select nodes with  degree meeting the following criteria:  in the sparse case, nodes with  degree 2, 3, and 4; in dense  networks, the degrees are 4, 5, and 6 for low, medium, and  high  data-complexity conditions respectively. For task 3 on the LobeHub task, we select lobes that contain at least  one hub  node.  For task  4, we first identify all hubs for each  network and  all neighbors of these  hubs  that  are located  in  the  opposite hemisphere of the  brain.  Of these neighboring nodes, we select  those  with  degree within the ranges [2,3], [4,5], and  [5,7] in the sparse condition and  [4,5] [6,7], and  [8,10] in the  dense  conditions for  low,  medium, and  high  data-complexity conditions, respectively.

\subsection{Hypotheses}

We have the following working hypotheses.

\begin{itemize}
\item{H1. For change-detection tasks, we would not observe differences in correctness among three layout approaches.}
\end{itemize}

This  is because these  three  methods lay  out  nodes at about  the  same  spatial proximity. 
%We  call  this  hypothesis the visual  equivalence hypothesis.

\begin{itemize}
\item{
H2. Circular and matrix would be more accurate than projection in the LobeHub tasks, but circular and projection would be more accurate than matrix in NeighborHub and HemisphereHub tasks.
}
\end{itemize}

This is because circular would support tasks  that  require hierarchy and  symmetry reading; projection supports 
symmetry but would be slow in hierarchy; matrix  supports better hierarchy but  perhaps not  symmetry. 
We believe this positioning would have influenced users for interpreting quantitative data  visualizations 
in network structures.

\begin{itemize}
\item{H3. 
For  change-detection tasks,  the  ranking  of the  visual markers would partially  
follow the Mackinlay order for quantitative  data comparison of length, angle, slope, area, density, lightness, and hue. 
}
\end{itemize}

We  think  this  ranking hypothesis would be  supported because this task is a simple  comparison of two quantitative values.

\begin{itemize}
\item{H4. For multi-scale comparisons, the rankings for hue, texture, and shape would improve, 
though they were ranked the last three in the Mackinlay order for showing quantities.
}
\end{itemize}

This  ranking hypothesis about  hue,  texture, and  shape would be supported because of our design choices. 
The hues  are  monotonic, texture   design  combines  size   encoding,  and   curvatures  in shapes are  
pre-attentive and thus  support efficient  visual  comparison. The  presence of these new features in the visual  
markers makes  us  to  believe  it is important to rank the marker effectiveness.

\begin{itemize}
\item{
H5. Length and area in general would lead to the most accurate answers and would consume the least amount of time.
}
\end{itemize}

This is because length  and  area are generally easier compared and  would be the more  salient  than color and  
orientation~\cite{garlandini2009evaluating}.

%to
%circular view (where rotation is needed) and projections (where the common spatially-aligned 0-axis does not exist.)

\subsection{Study Design}

We  use  a full-factorial design with  two  independent variables: layout (three levels) and  visual  marker (nine levels)  
(Table~\ref{tab:expDesign}). We also  selected three  levels  of  data  complexity (low,  medium, and  high) and  two  levels of edge  density (sparse and  dense) in order to include a broad range of data.  
The three  layout methods are combined with the three  data complexities in a
 $3\times3$ Latin square.
Participants performed one of the three
squares using  the nine markers with  both  sparse and  dense edge  densities. Thus,  each participant performed 
$3$ (layout-data complexity) $\times$ 9 (markers) $\times$ 2 (edge density)=54 trials in each  task  and  a total  of 
216 trials  for all four  tasks.  The 54  trials   in  each  task   were   randomly  ordered  to  avoid learning effects. 
All participants completed all trials  in task 1 followed by all trials in task types  2, 3, and  4 in that order.

\begin{table}
	%% Table captions on top in journal version
	\caption{
Experimental Design: each participant performed 54 trials for each of the 4 task types. Here layout
and data complexity follow a Latin square combination; this combination and marker 
form a  within-participant design that each participant looks all conditions. 
The orders of marker type are randomized when combined with three layout-complexity combinations
for each participant.
	}
	\label{tab:expDesign}
%	\scriptsize
	\begin{center}
		\begin{tabular}{l | l | l | l}
		\hline
			Participant & Layout-               & Marker & Edge \\ 
			 ID                  & data complexity  & type &  density\\
			\hline
			     & Projection-low  & & \\
			1-6 &  Circular-medium & & \\
			&  Matrix-high &   & Sparse ($5\%$) \\
			\cline{1-1}\cline{2-1} 
			   & Circular-low  &   9 types &  and\\
			7-12 &  Matrix-medium & & dense ($10\%$) \\
			&  Projection-high & &  \\
			\cline{1-1}\cline{2-1}
			 & Matrix-low   &   &     \\
			13-18 & Projection-medium & & \\
			&  Circular-high & & \\
			\hline
		\end{tabular}
	\end{center}
\end{table}

%\subsection{Task Data Generation}
%The degree distribution in the filtered graphs follows the distribution of brain networks in Achard et al.~\cite{achard2006resilient}.

\subsection{Participants}

Eighteen participants volunteered for  the  study and received minimum wage  compensation. Their  ages ranged  from  18 to 29 with  average 22.4 (standard deviation=3.0); eight are female.  Seven studied computer science (4 females),  one computer engineering, two  mechanical engineering, two  human-computer  interaction (one  female),  one  psychology (female),  one  in  biology (one female),  and one  Asian  studies (female).  The three  brain scientists participated in the study and were 
evenly  distributed in these three  participant ID groups. Females were also distributed in these 
three  groups as evenly as possible.

\subsection{Procedure}

The experiment was  conducted in a quiet  lab. The display was  a  Dell  $23''$  monitor; 
the  screen   resolution was
$1,920\times1,200$. Participants first completed a consent form and then  a background survey. Three  layout and  nine  markers 
were  introduced. The training session  ensured that  the participants had  fully understood all encoding approaches and 
the task  requirements. They  were  given  nine  practice tasks for  each  task  type.  Answers were  shown on  the  screen  
in the  training session  to  let  the  participants learn  by  fixing
their mistakes. The  training session  were  not timed  and  participants were  instructed to fully understand the  tasks  and  
encoding methods before  proceeding to  the formal  testing, when participants completed tasks  independently and  were  
allowed to quit  the  study or  take  breaks at  any  time.  
During the formal testing, our  program logged   task  completion  time, all interaction, and  
participants' answers. 
Participants completed  a post-questionnaire to  rate  the  techniques and  were interviewed for comments.

\section{Results}

This section  presents statistical analysis results.

\begin{table}[t!]
%% Table captions on top in journal version
 \caption{Summary Statistics by Tasks. ES: Effect size}
 \label{tab:expResults}
 \scriptsize
 \centering
   \begin{tabular}{c | l | l | l }
   \hline
     Task & Variable & Statistical test & ES \\
   \hline 
               &  Accuracy (marker)    &\textbf{$\chi^2_{(8, 972)}$=23.3, p=0.003}  &\textbf{0.15}\\
Change-          &  Accuracy (layout)            &  $\chi^2_{(2, 972)}=0.3, p=2.6$                 & 0.05\\ 
detection	  &  Time (marker)              & $F_{(8, 948)}=1.2, p=0.27$   & 0.30 \\
               & Time (layout)      & \textbf{$F_{(2, 948)}=16.0$}, p$<$0.0001                & \textbf{0.44}\\ 
         %   &  \textbf{Time}         &  \textbf{F(16, 948)=1.8}               & \textbf{d=1.15}\\ 
          %   & (\textbf{marker $\times$ layout}) & \textbf{p=0.03} & \\
    \hline
  
               & Accuracy (marker)        &  \textbf{$\chi^2_{(8, 972)}=36.9$, p$<$0.0001}                  &\textbf{0.20}\\ 
   Neighbor  & Accuracy (layout)    & \textbf{$\chi^2_{(2, 972)}=9.1$, p=0.01}                 & \textbf{0.10}\\ 
          Hub      & Time (marker)       & \textbf{$F_{(8, 950)}=6.3$, p$<$0.0001}                & \textbf{0.66}\\
               & Time (layout)          & \textbf{$F_{(2, 950)}=317.7$, p$<$0.0001}                & \textbf{1.34}  \\ 
         %      & \textbf{Time} \textcolor{red}{NEW}     & \textbf{F(16, 950)=2.0} & \textbf{d=1.89}\\
          %      & \textbf{(marker $\times$ layout)} & \textbf{p=0.01} & \\ 
               \hline

               &  Accuracy (marker)       & \textbf{$\chi^2_{(2, 972)}=36.7$, p$<$0.0001}                &\textbf{0.19}\\ 
 Lobe  & Accuracy (layout)            & \textbf{$\chi^2_{(2, 972)}=41.8$, p$<$0.0001}                 &\textbf{0.21}\\
         Hub       &  Time (marker)  & \textbf{$F_{(8, 952)}=14.8$, p$<$0.0001}   	        & \textbf{1.13} \\
               &  Time (layout)       &  \textbf{$F_{(2, 952)}=28.6$, p$<$0.0001}                &  \textbf{0.55} \\ 
       %       &  \textbf{Time}         &  \textbf{F(16, 952)=2.1}              & \textbf{d=1.86}\\
        %       & \textbf{(marker $\times$ layout)} & \textbf{p=0.009} & \\ 
    \hline 
    
               &  Accuracy (marker)       & \textbf{$\chi^2_{(2, 972)}=18.3$, p=0.02}              &\textbf{0.14}\\  
Hemisphere             &  Accuracy (layout)         & \textbf{$\chi^2_{(8, 972)}=27.7$, p$<$0.0001}                 &\textbf{0.17}\\
  	 Hub       &  Time (marker)  & \textbf{$F_{(8, 952)}=3.2$, p=0.001}   	        & \textbf{0.39} \\
               &  Time (layout)   & \textbf{$F_{(2, 952)}=335.1$, p$<$0.0001}               & \textbf{1.40} \\ 
      %     &  \textbf{Time}         &  \textbf{F(16, 952)=2.5}             &\textbf{d=2.02}\\ 
      %        & \textbf{(marker $\times$ layout)} & \textbf{p=0.001} & \\
               \hline   
    \end{tabular}
\end{table}

\subsection{Analysis Approaches}

%Table~\ref{tab:expResults} shows the $F$ and $p$ values computed with general linear model (GLM) procedure on time. Tukey pairwise comparisons among dependent valuables of encoding and layout are computed in post-hoc analysis on time. The correctness data were binary (correct or incorrect) and was analyzed using logistic regression. The \textit{p} value for Wald Chi-Square statistic test are reported. When the \textit{p} value is less than 0.05, variable levels with 95\% confidence interval of pair-wise difference of odds ratios not overlapping 1.0 are considered significantly different. All figures use $95\%$ confidence interval in the error bar. SAS's  test with the ``freq'' procedure is used to examine whether or not there is a significant correlation between the main effect (either encoding or layout) and the accuracy.

We collected  3,888 data  points (972 from  each  of the  four  tasks)  and  analyzed the  results by  task  using  the statistics 
analysis software SAS. $F$ and  $p$ values of the main effects  and  their  interaction with  task  completion 
time  are computed with  a general linear  model  (GLM) procedure. For significant main  effects,  a  post-hoc analysis 
with  Tukey's honest significant difference (HSD) test is used.  The correctness data  are binary (correct  or incorrect) and  are analyzed using   logistic  regression and  reported using  the  $p$  value from  the  Wald   $\chi^2$ test.  
When  the   $p$ value  is less than  0.05, variable levels  with $95\%$ confidence interval  of  
pairwise difference  of  odds ratios  not  overlapping are  considered significantly different. The 
 $\chi^2$ test  with  the ``freq''  procedure is  used   to  examine whether or  not  there  is  
 a  significant correlation between  the  main   effect  (either positioning or marker) and  accuracy.

We   measure  effect   sizes   using   Cohen's  $d$  for   time
and  Cramer's $V$   for  correctness to  understand the  practical significance~\cite{cohen1988statistical}. 
We used  Cohen's  benchmarks for ``small''(0.07-0.21,) ``medium'' (0.21-0.35,) and ``large'' ($>0.35$) effects.  
We  separate our  analyses by  task and  give
the summary statistics in Table{~\ref{tab:expResults}} and Figs.~\ref{fig:task1}-\ref{fig:task4}.

\begin{comment}
%\subsection{Overview}
%Table~\ref{tab:expResults} shows the summary statistics of the main effects of encoding and layout on task completion time and
%accuracy. Figure~\ref{fig:overall} shows the aggregated performance. 
%Of the three layout techniques, circular view leads to the most accurate answer and takes the participants the least amount of 
%time to complete. 
%Of the nine encoding approaches, area, length, and angle in general have the most accurate answers and the shortest task 
%completion time. 
%Hue brings relatively high accuracy but is also time demanding. 
%Density, texture, and shape are similar; lightness both is slow and causes many errors. 
%Slope was took the longest task completion time with the least accurate answers.

%\input{ResultSummaryFigures2.tex}
\end{comment}

\subsection{Summary Statistics by Tasks}

\begin{figure}[t!]
\centering
\begin{subfigure}{.5\linewidth}
	\centering`
	\includegraphics[width=\linewidth ]{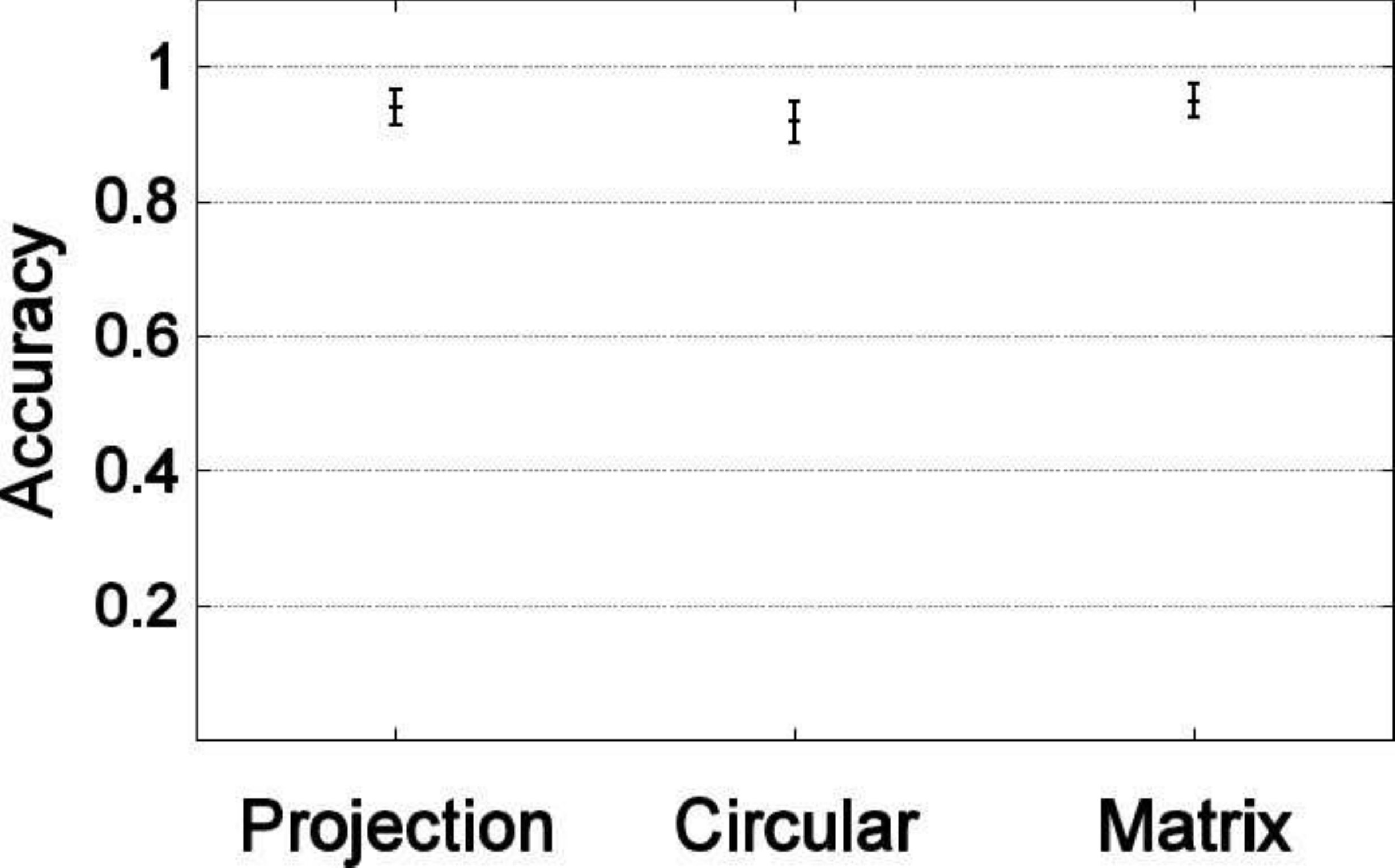}
	\caption{Layout vs. Accuracy}
	\label{fig:accLayoutTask1}
\end{subfigure}%
\begin{subfigure}{.5\linewidth}
	\centering
	\includegraphics[width=\linewidth ]{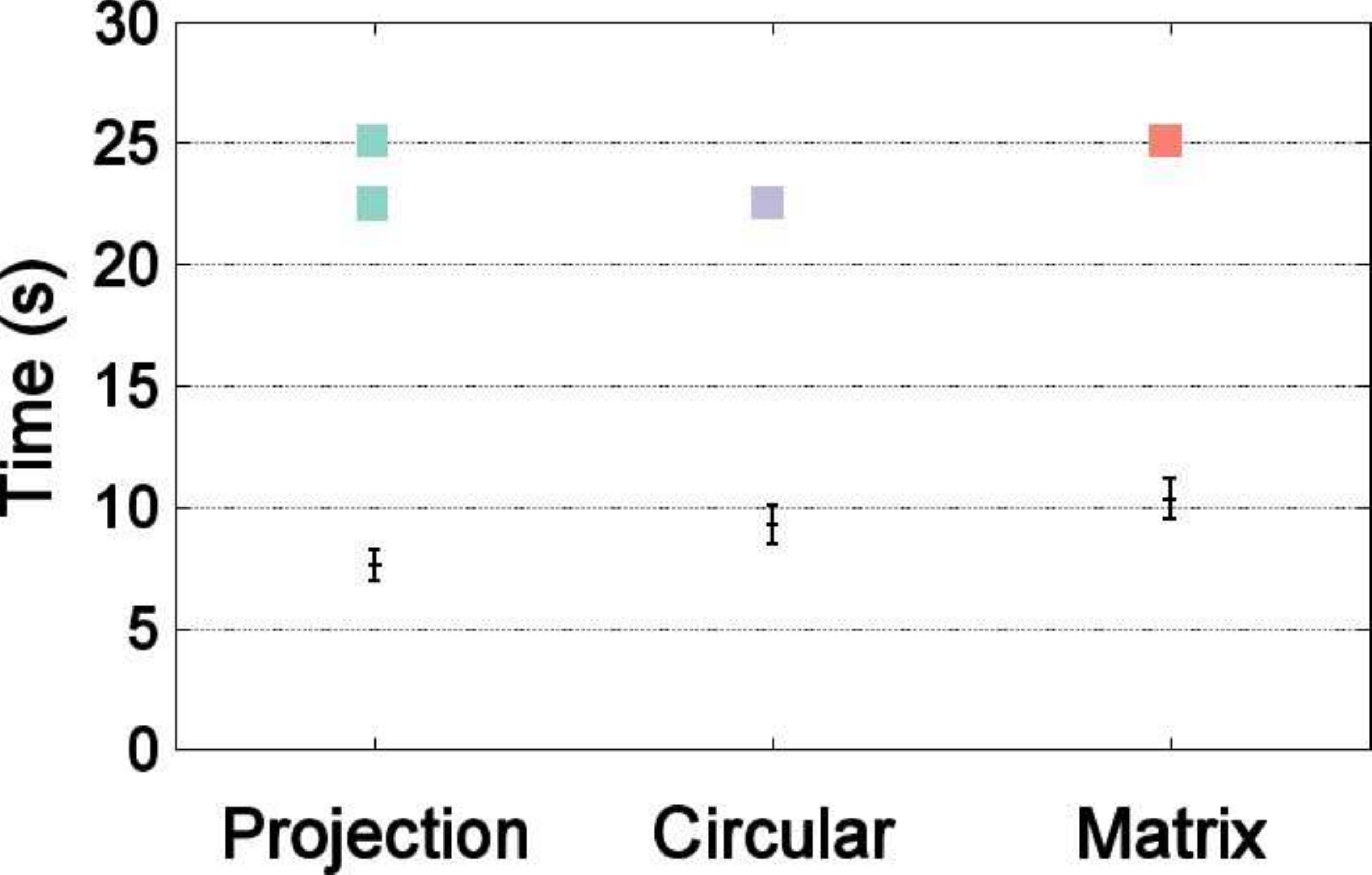}
	\caption{Layout vs. Time}
	\label{fig:timeLayoutTask1}
\end{subfigure}%

\begin{subfigure}{.5\linewidth}
	\centering
       \includegraphics[width=\linewidth ]{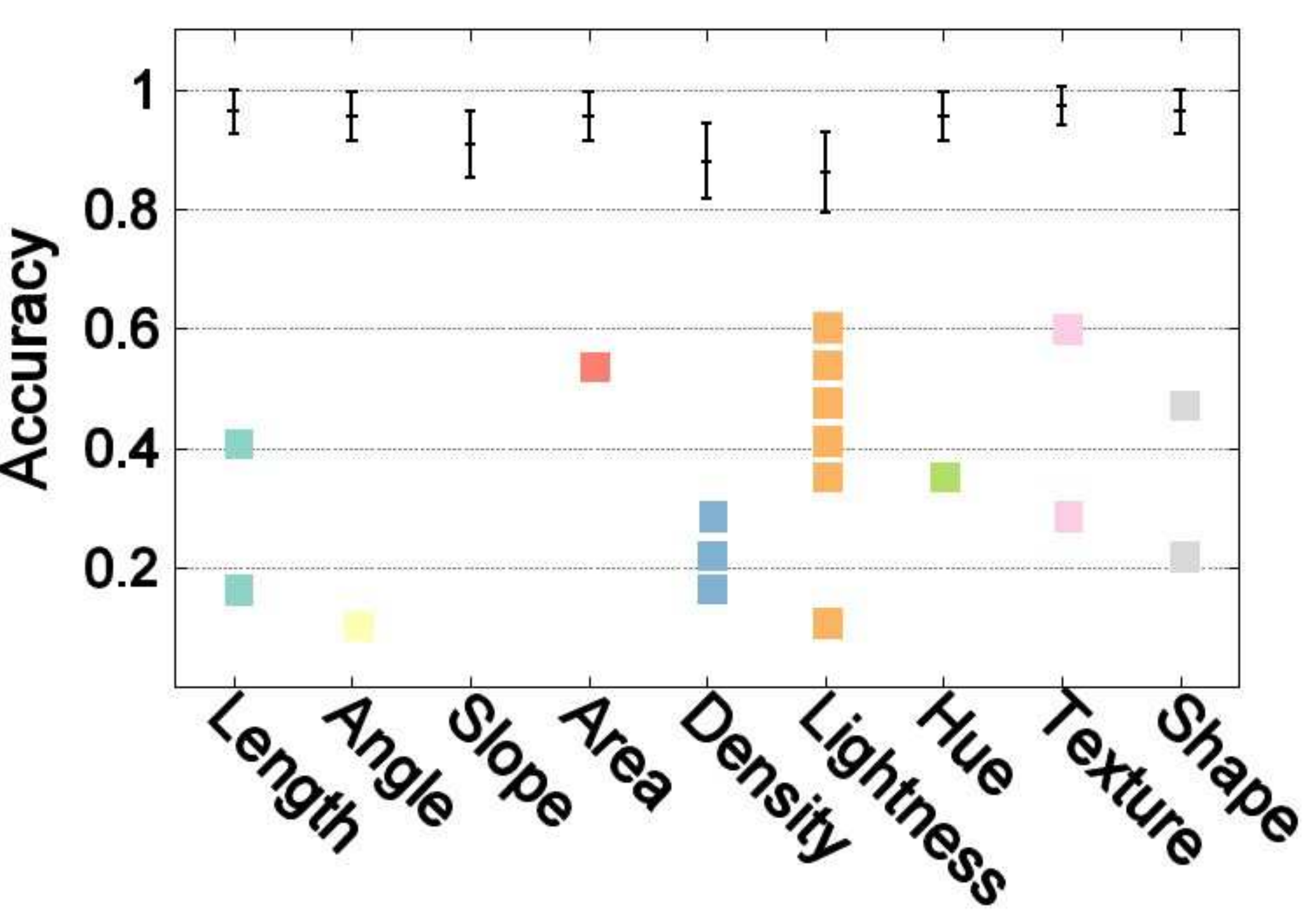}
	\caption{Encoding vs. Accuracy}
	\label{fig:accEncodingTask1}
\end{subfigure}%
\begin{subfigure}{.5\linewidth}
	\centering
       \includegraphics[width=\linewidth ]{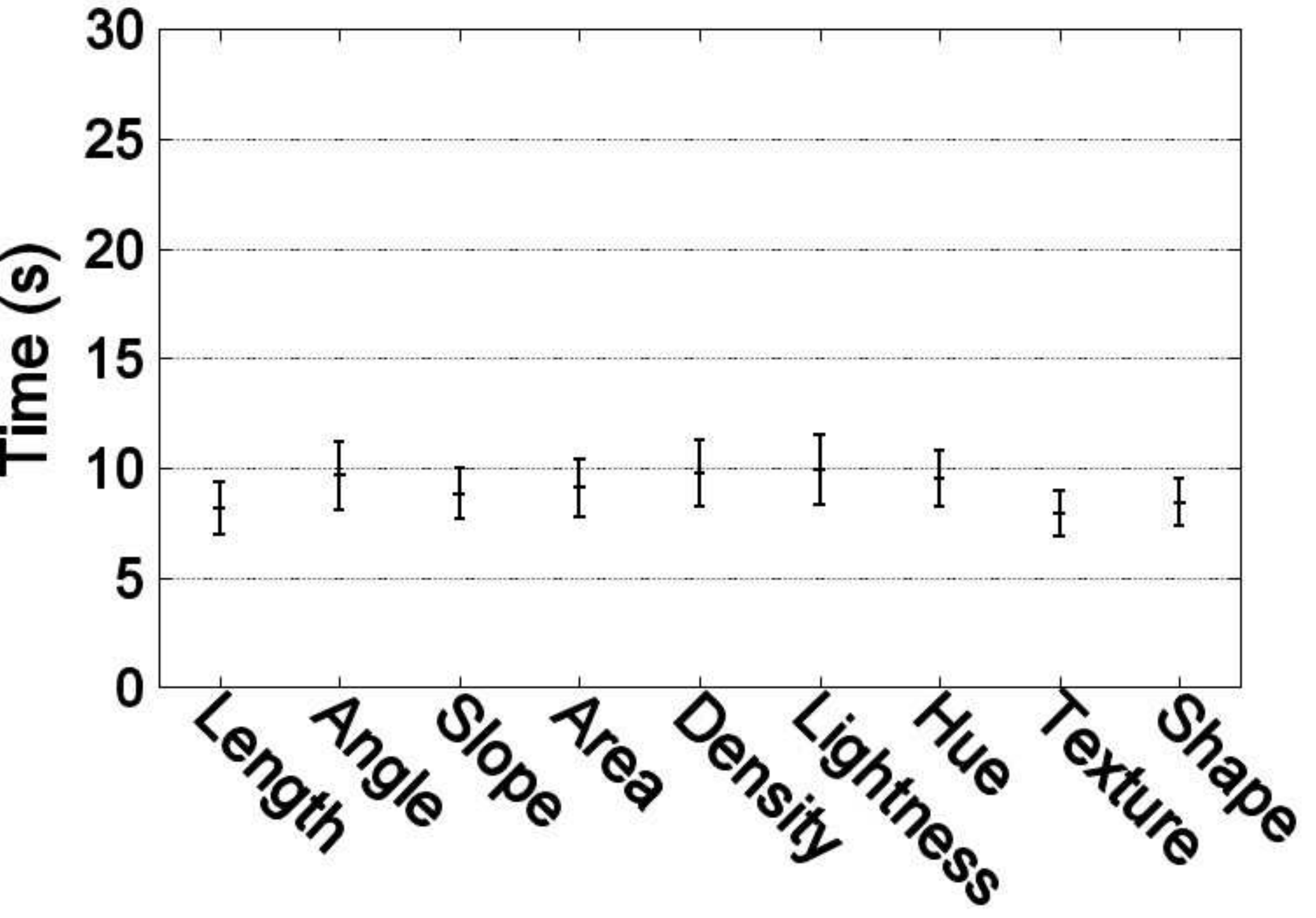}
	\caption{Marker Types vs. Time}
	\label{fig:timeEncodingTask1}
\end{subfigure}%

\begin{comment}
\begin{subfigure}{.5\linewidth}
	\centering
%	\includegraphics[width=\linewidth ]{images/expResult/AccTask1.pdf}
\includegraphics[width=\linewidth ]{images/interaction_correctness_t1.pdf}
	\caption{Marker Types by Layout vs. Accuracy}
	\label{fig:accIntTask1}
\end{subfigure}%
\begin{subfigure}{.5\linewidth}
	\centering
%	\includegraphics[width=\linewidth ]{images/expResult/TimeTask1.pdf}
\includegraphics[width=\linewidth ]{images/interaction_time_t1.pdf}
	\caption{Marker Types by Layout vs. Time}
	\label{fig:timeIntTask1}
\end{subfigure}%
\end{comment}

\caption{Task type 1: Change-detection}
\label{fig:task1}
\end{figure}

\begin{figure}[t]
\centering
\begin{subfigure}{.5\linewidth}
	\centering
	\includegraphics[width=\linewidth ]{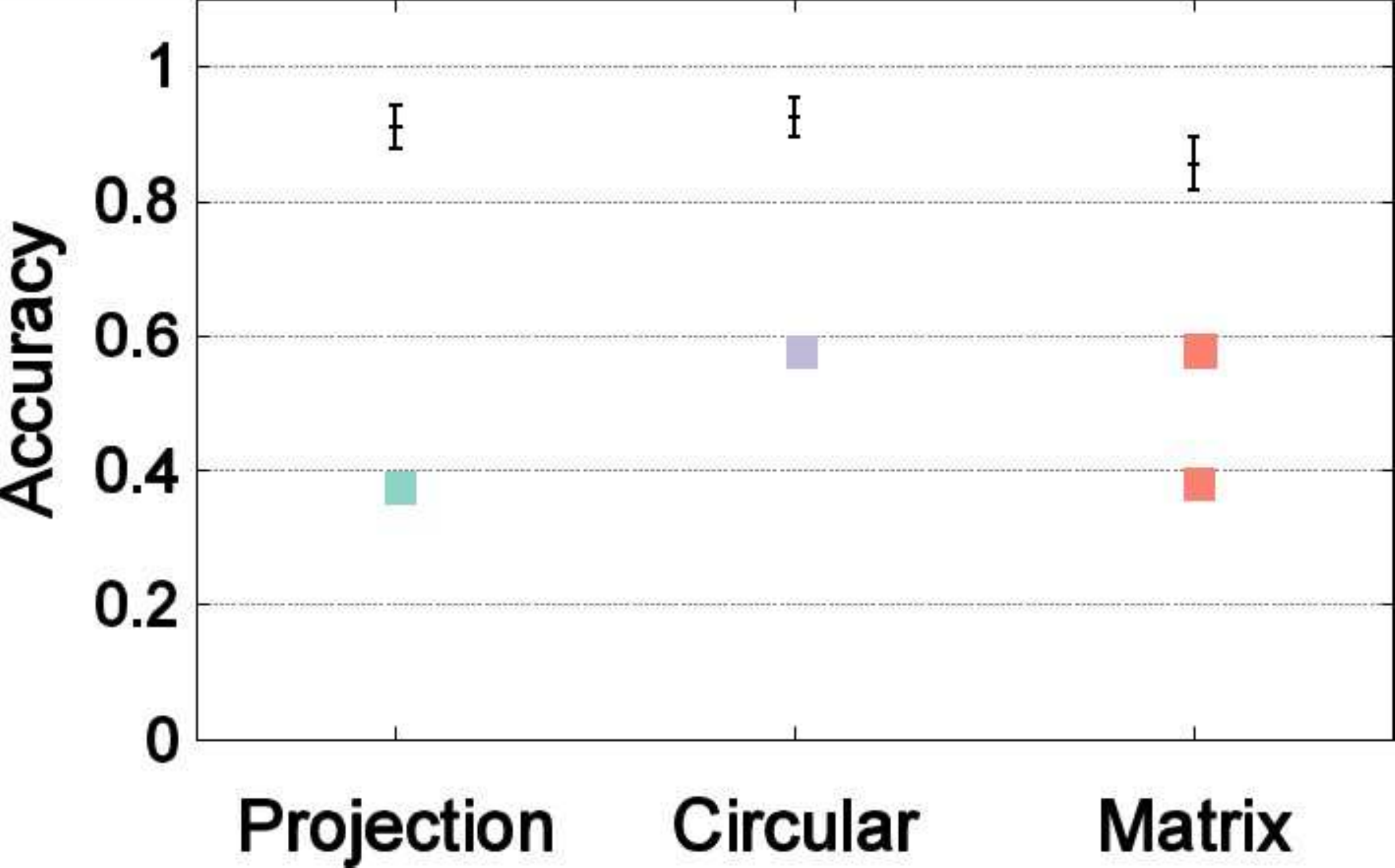}
	\caption{Layout vs. Accuracy}
	\label{fig:accLayoutTask2}
\end{subfigure}%
\begin{subfigure}{.5\linewidth}
	\centering
	\includegraphics[width=\linewidth ]{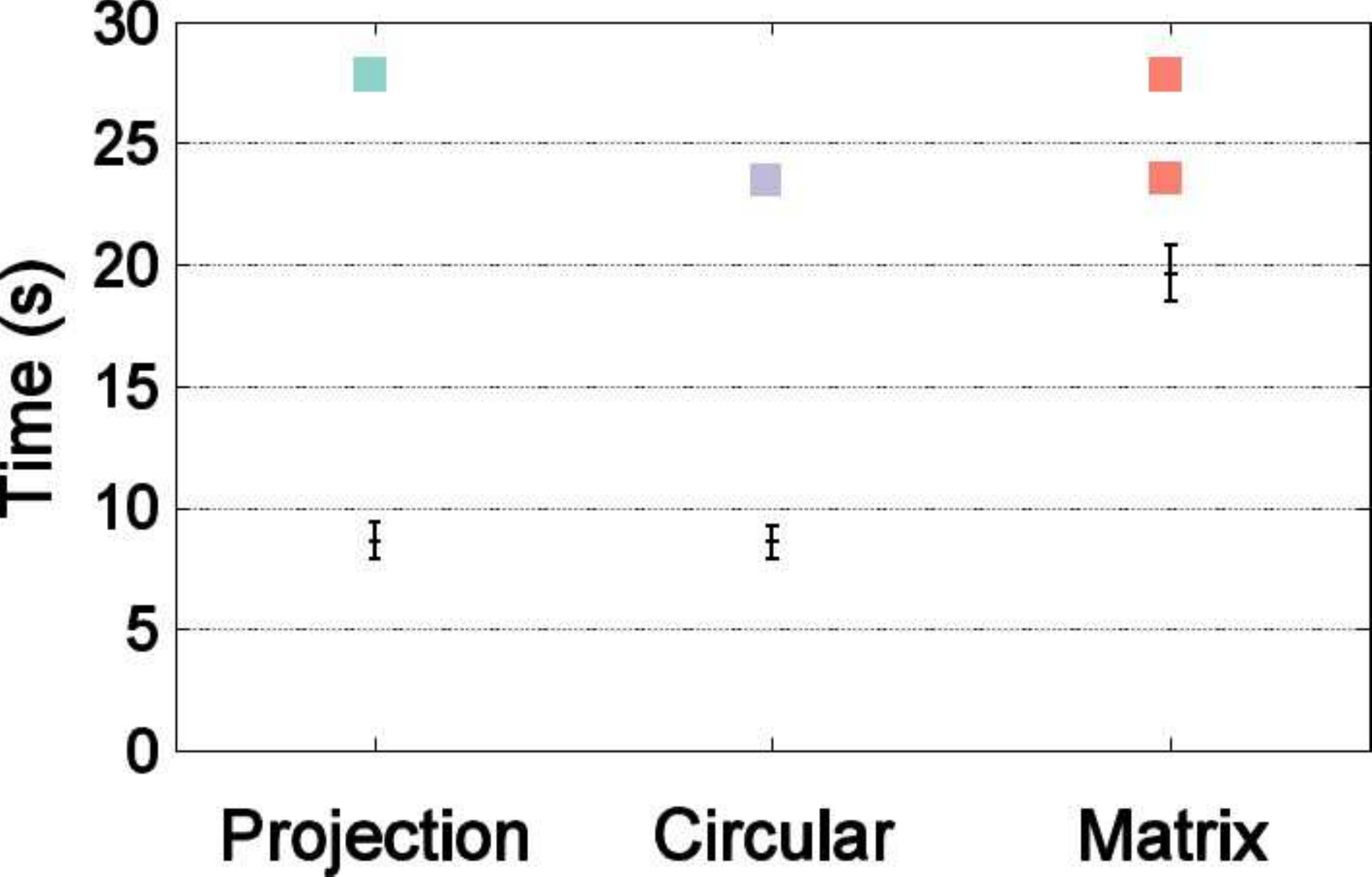}
	\caption{Layout vs. Time}
	\label{fig:timeLayoutTask2}
\end{subfigure}%

\begin{subfigure}{.5\linewidth}
	\centering
       \includegraphics[width=\linewidth ]{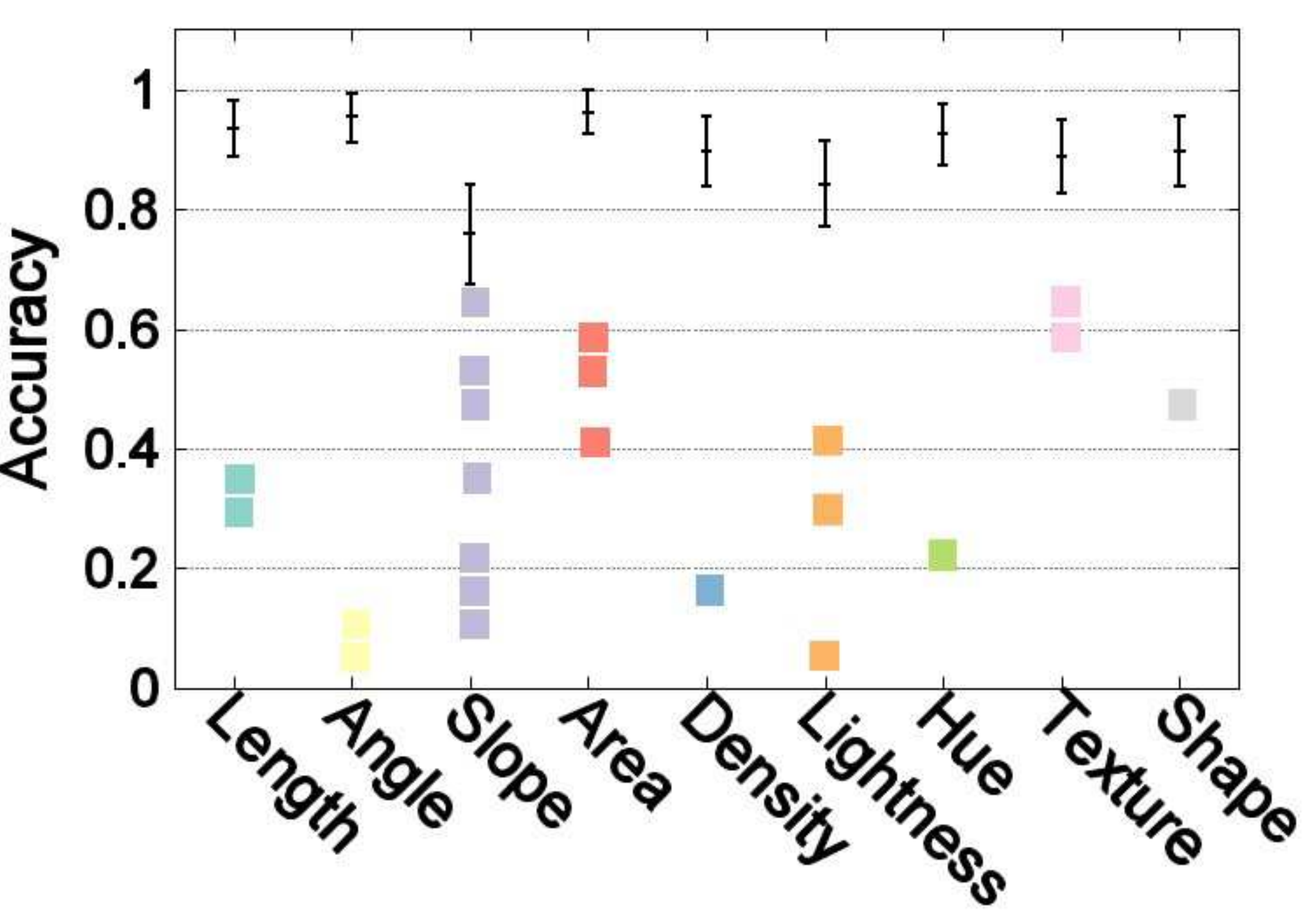}
	\caption{Encoding vs. Accuracy}
	\label{fig:accEncodingTask2}
\end{subfigure}%
\begin{subfigure}{.5\linewidth}
	\centering
       \includegraphics[width=\linewidth ]{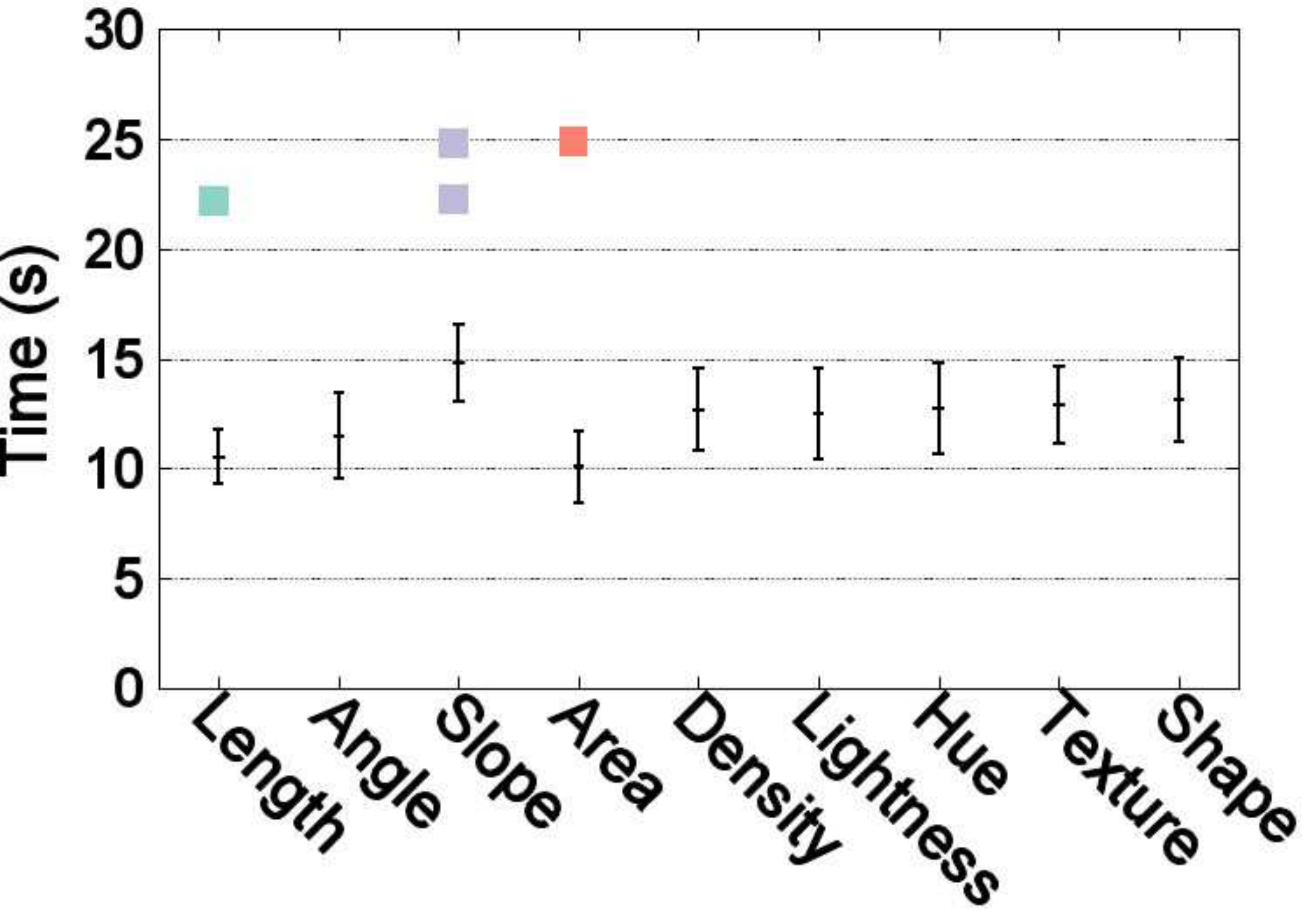}
	\caption{Marker Types vs. Time}
	\label{fig:timeEncodingTask2}
\end{subfigure}%

\begin{comment}
\begin{subfigure}{.5\linewidth}
	\centering
%	\includegraphics[width=\linewidth ]{images/expResult/AccTask1.pdf}
\includegraphics[width=\linewidth ]{images/interaction_correctness_t2.pdf}
	\caption{Marker Types by Layout vs. Accuracy}
	\label{fig:accIntTask2}
\end{subfigure}%
\begin{subfigure}{.5\linewidth}
	\centering
%	\includegraphics[width=\linewidth ]{images/expResult/TimeTask1.pdf}
\includegraphics[width=\linewidth ]{images/interaction_time_t2.pdf}
	\caption{Marker Types by Layout vs. Time}
	\label{fig:timeIntTask2}
\end{subfigure}%
\end{comment}

\caption{Task type 2: NeighborHub}
\label{fig:task2}
\end{figure}

\begin{figure}[t]
\centering
\begin{subfigure}{.5\linewidth}
	\centering
	\includegraphics[width=\linewidth ]{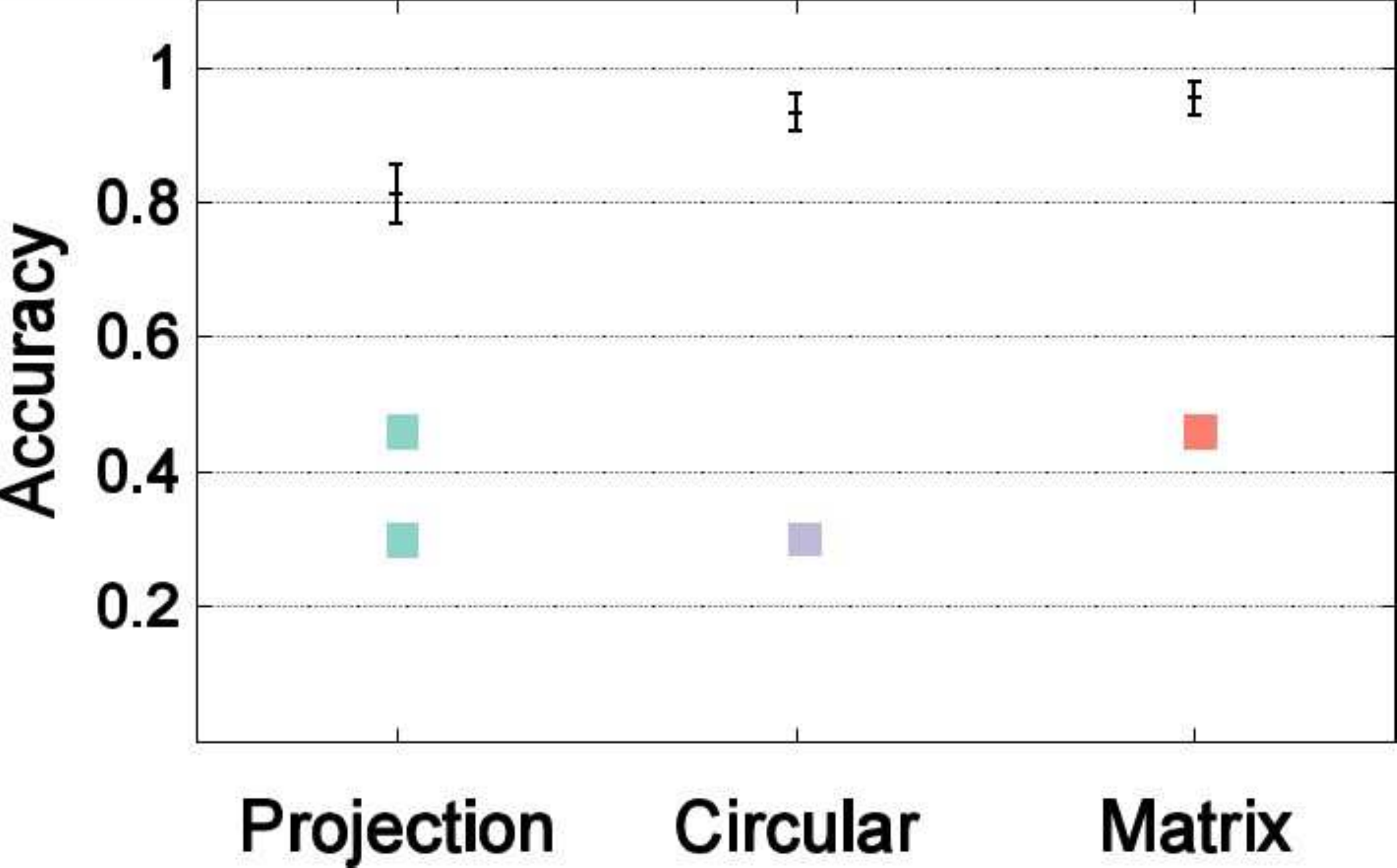}
	\caption{Layout vs. Accuracy}
	\label{fig:accLayoutTask3}
\end{subfigure}%
\begin{subfigure}{.5\linewidth}
	\centering
	\includegraphics[width=\linewidth ]{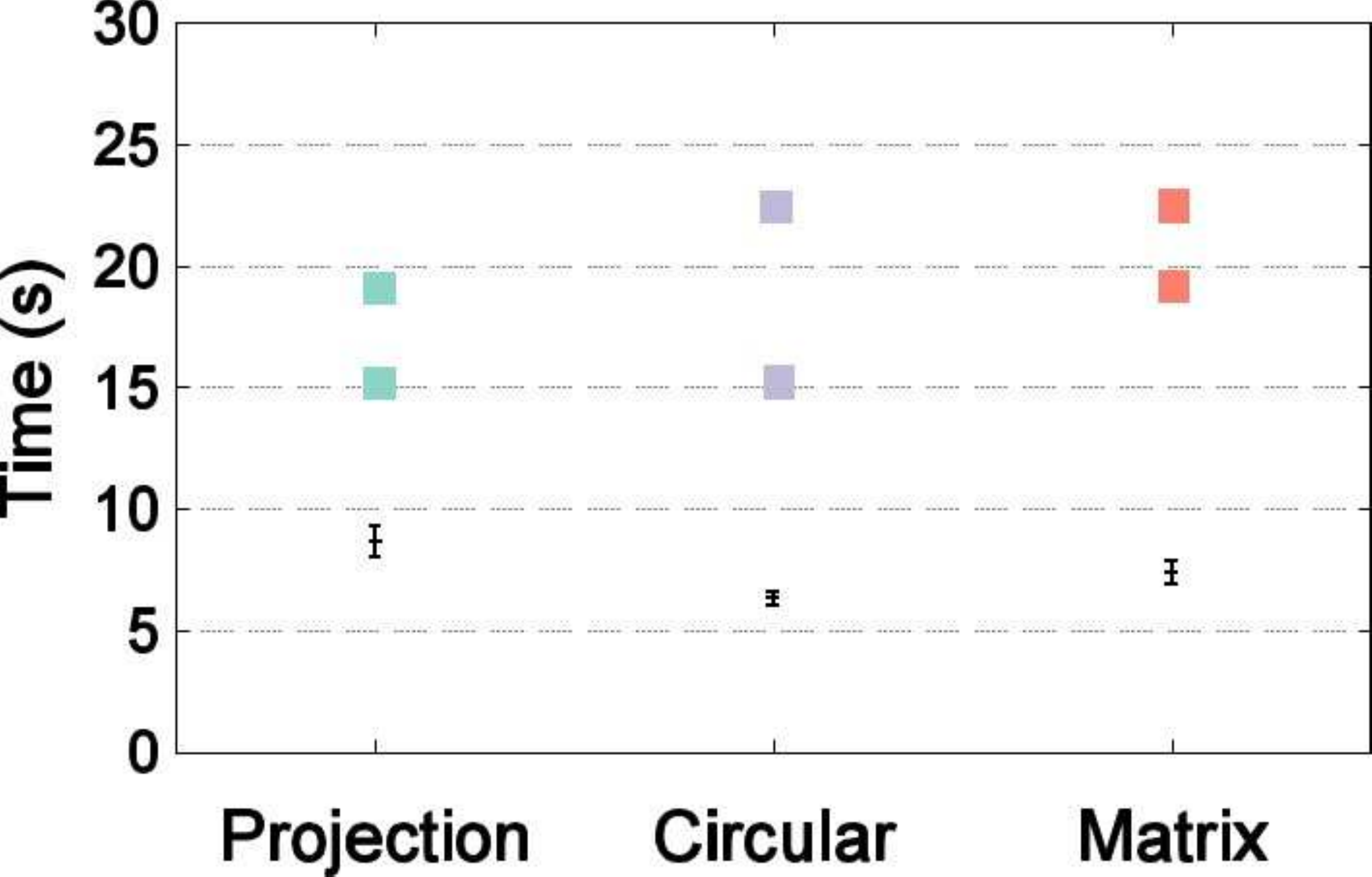}
	\caption{Layout vs. Time}
	\label{fig:timeLayoutTask3}
\end{subfigure}%

\begin{subfigure}{.5\linewidth}
	\centering
       \includegraphics[width=\linewidth ]{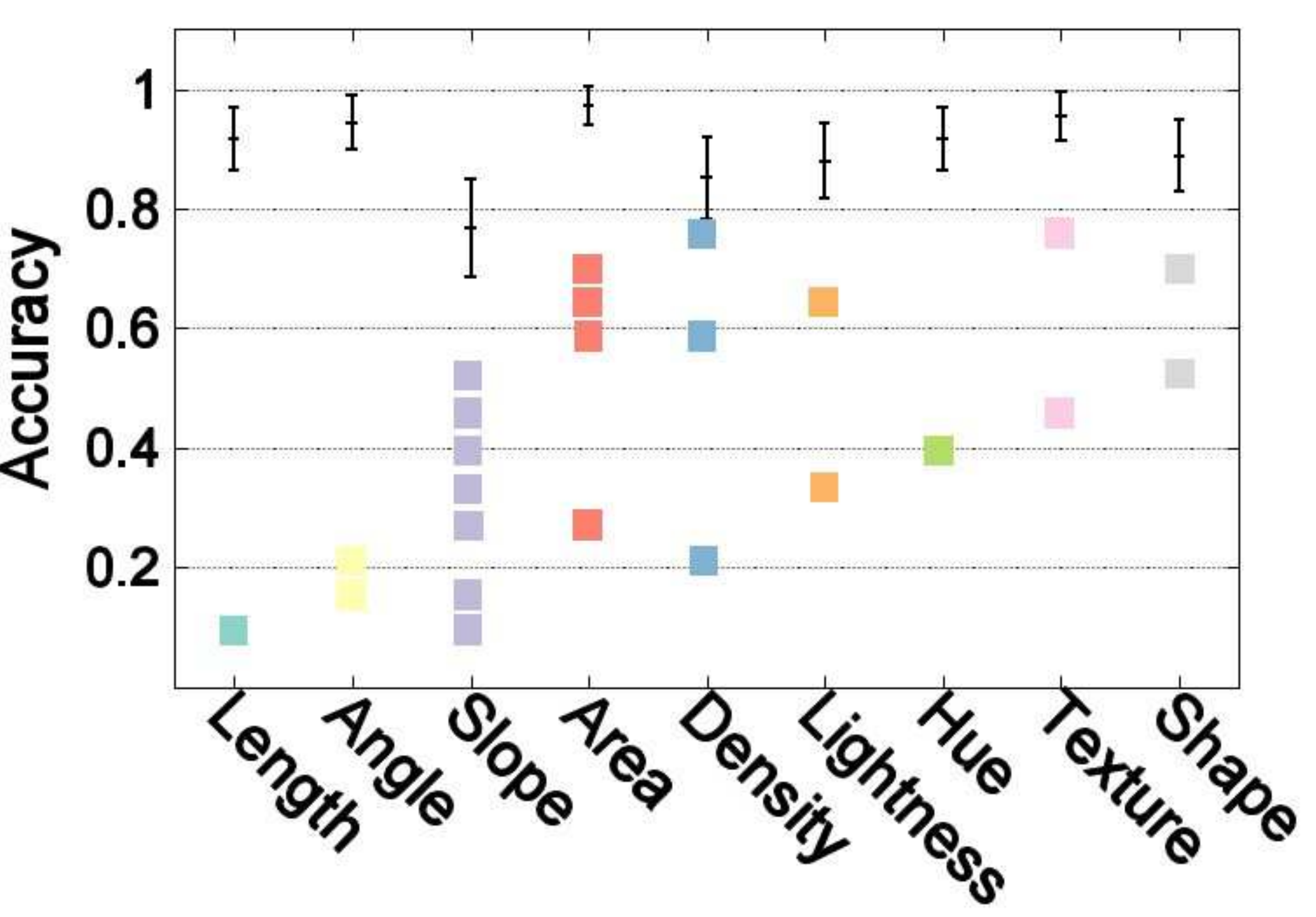}
	\caption{Encoding vs. Accuracy}
	\label{fig:accEncodingTask3}
\end{subfigure}%
\begin{subfigure}{.5\linewidth}
	\centering
       \includegraphics[width=\linewidth ]{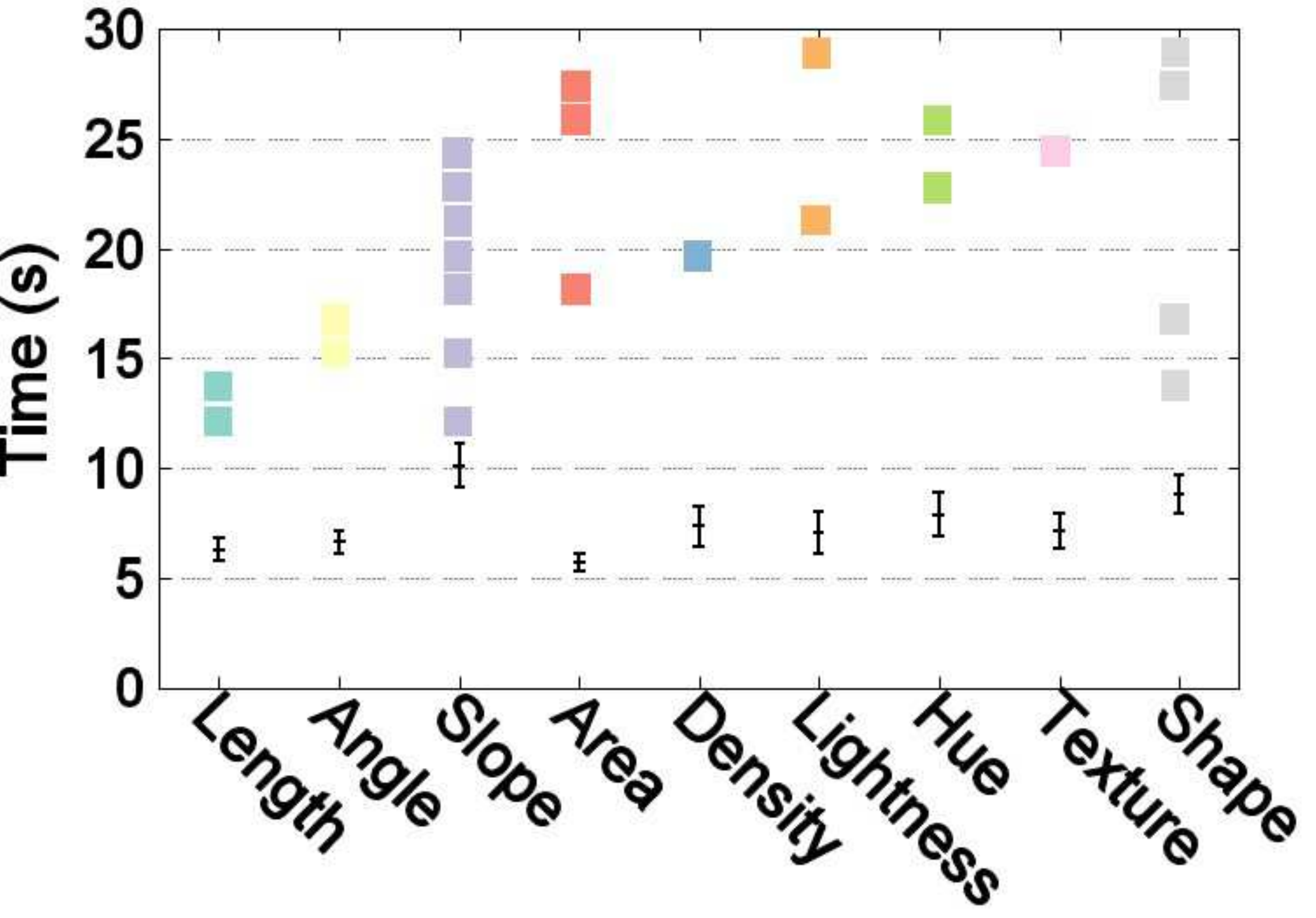}
	\caption{Marker Types vs. Time}
	\label{fig:timeEncodingTask3}
\end{subfigure}%

\begin{comment}
\begin{subfigure}{.5\linewidth}
	\centering
%	\includegraphics[width=\linewidth ]{images/expResult/AccTask1.pdf}
\includegraphics[width=\linewidth ]{images/interaction_correctness_t3.pdf}
	\caption{Marker Types by Layout vs. Accuracy}
	\label{fig:accIntTask3}
\end{subfigure}%
\begin{subfigure}{.5\linewidth}
	\centering
%	\includegraphics[width=\linewidth ]{images/expResult/TimeTask1.pdf}
\includegraphics[width=\linewidth ]{images/interaction_time_t3.pdf}
	\caption{Marker Types by Layout vs. Time}
	\label{fig:timeIntTask3}
\end{subfigure}%
\end{comment}

\caption{Task type 3: LobeHub}
\label{fig:task3}
\end{figure}

\begin{figure}[t]
\centering
\begin{subfigure}{.5\linewidth}
	\centering
	\includegraphics[width=\linewidth ]{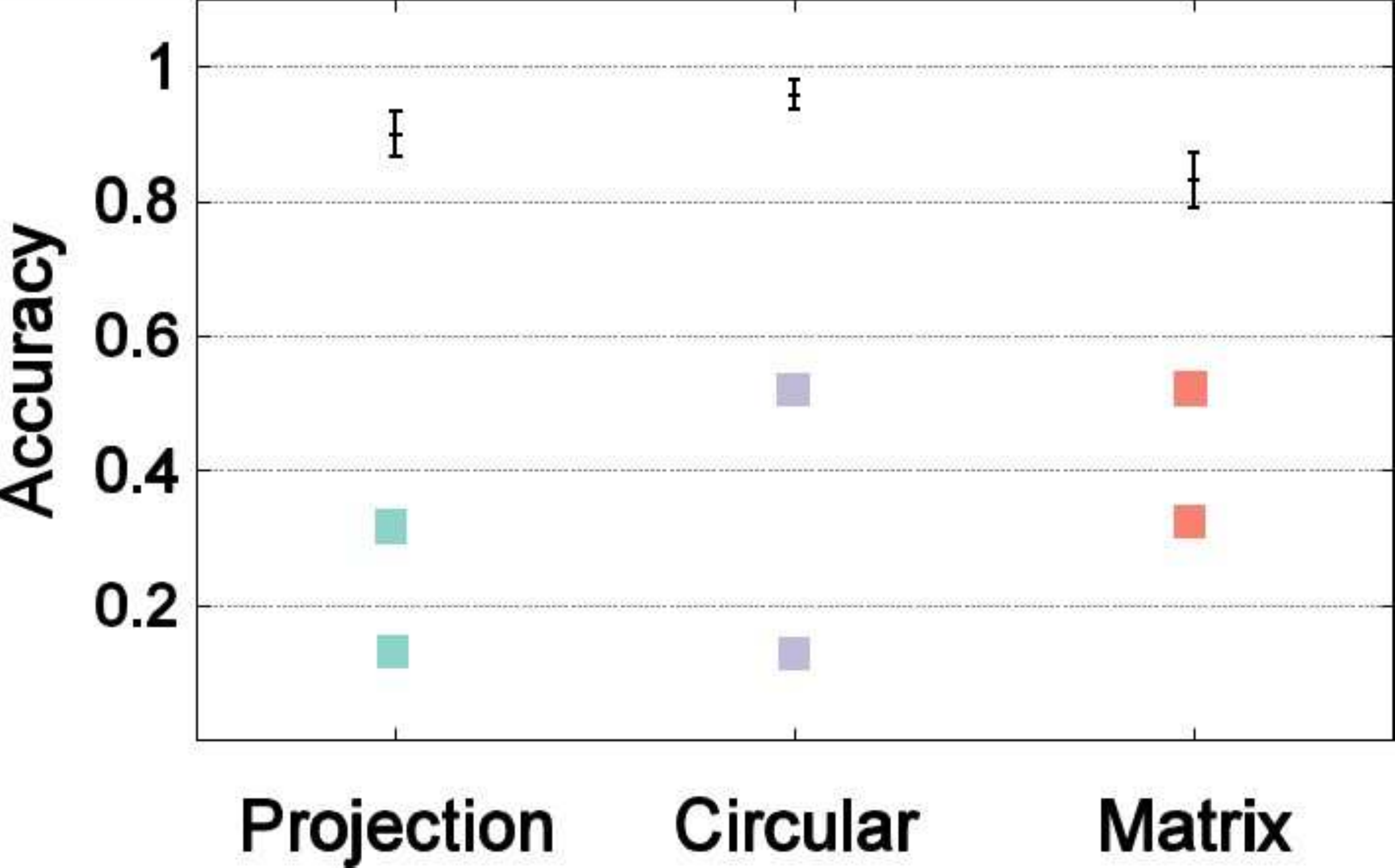}
	\caption{Layout vs. Accuracy}
	\label{fig:accLayoutTask4}
\end{subfigure}%
\begin{subfigure}{.5\linewidth}
	\centering
	\includegraphics[width=\linewidth ]{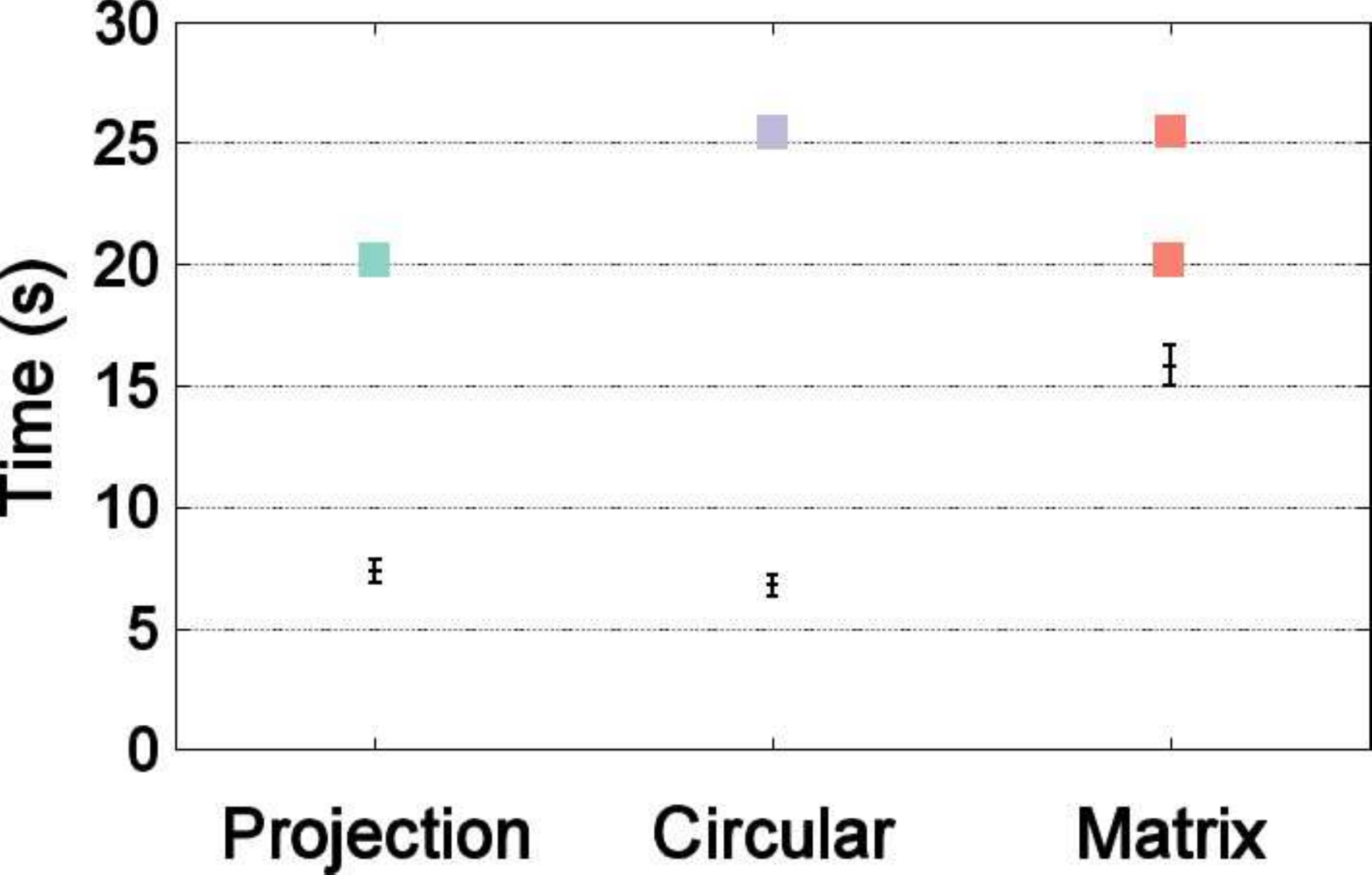}
	\caption{Layout vs. Time}
	\label{fig:timeLayoutTask4}
\end{subfigure}%

\begin{subfigure}{.5\linewidth}
	\centering
       \includegraphics[width=\linewidth ]{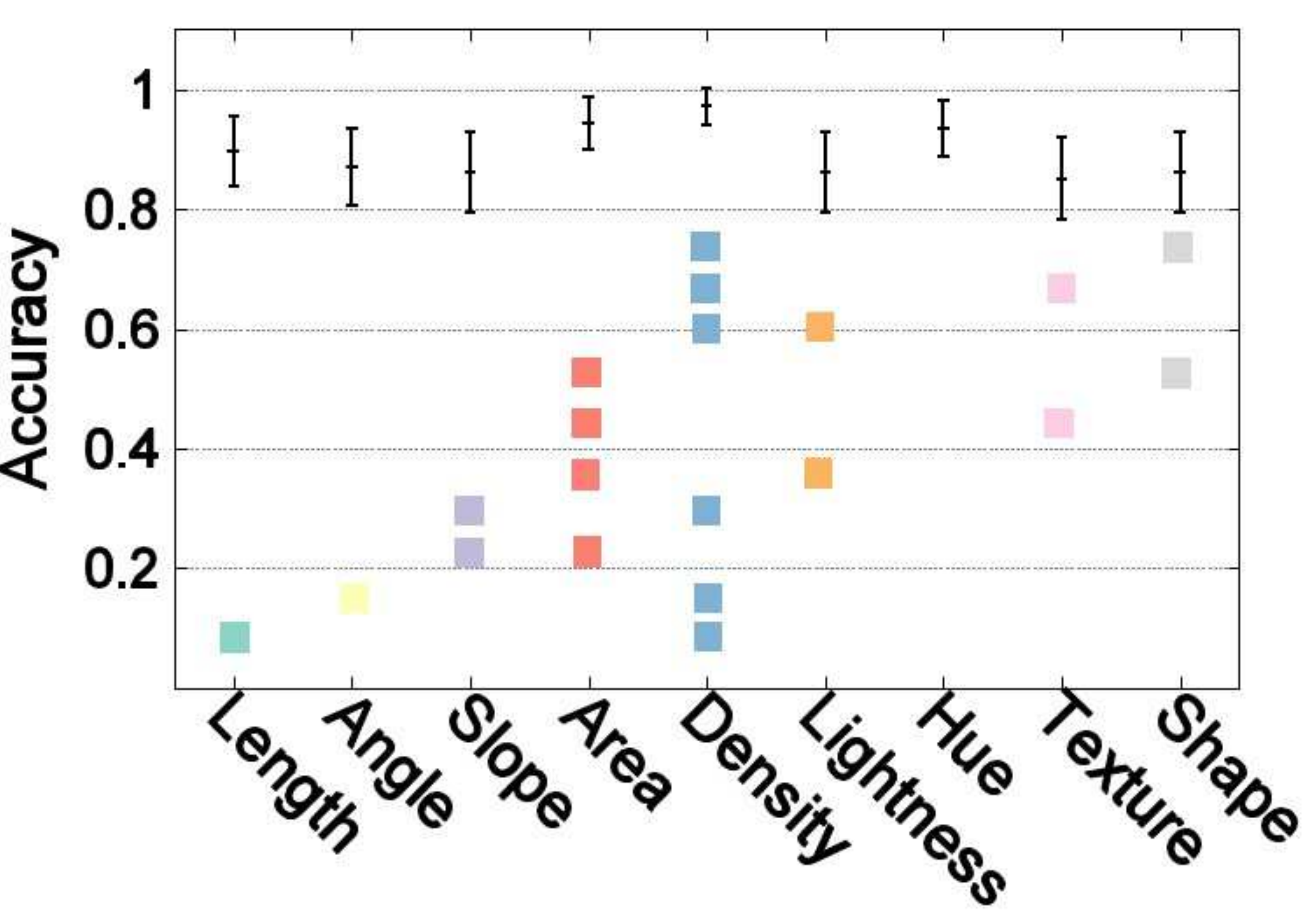}
	\caption{Marker Types vs. Accuracy}
	\label{fig:accEncodingTask4}
\end{subfigure}%
\begin{subfigure}{.5\linewidth}
	\centering
       \includegraphics[width=\linewidth ]{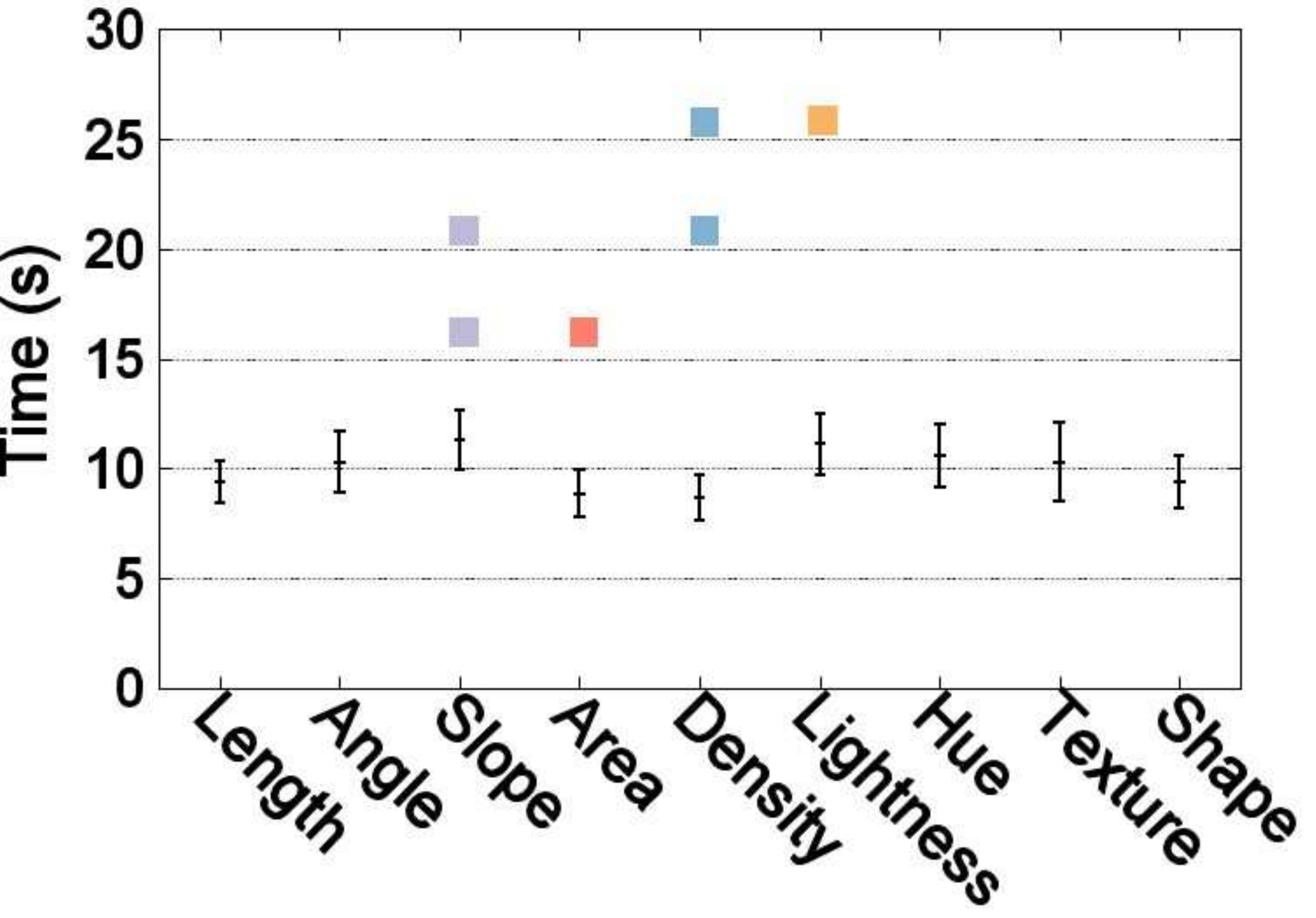}
	\caption{Marker Types vs. Time}
	\label{fig:timeEncodingTask4}
\end{subfigure}%

\begin{comment}
\begin{subfigure}{.5\linewidth}
	\centering
%	\includegraphics[width=\linewidth ]{images/expResult/AccTask1.pdf}
\includegraphics[width=\linewidth ]{images/interaction_correctness_t4.pdf}
	\caption{Marker Types by Layout vs. Accuracy}
	\label{fig:accIntTask4}
\end{subfigure}%
\begin{subfigure}{.5\linewidth}
	\centering
%	\includegraphics[width=\linewidth ]{images/expResult/TimeTask1.pdf}
\includegraphics[width=\linewidth ]{images/interaction_time_t4.pdf}
	\caption{Marker Types by Layout vs. Time}
	\label{fig:timeIntTask4}
\end{subfigure}%
\end{comment}

\caption{Task type 4: HemisphereHub}
\label{fig:task4}
\end{figure}

All subfigures in Figs.~\ref{fig:task1}-\ref{fig:task4} use  $95\%$ confidence intervals in the  error  bar.  The  colored  dots  along  the  same  horizontal line  indicate  significantly different  pairs   in  the  post-hoc analysis. For each task, we plot  the statistics 
for positioning and marker types  vs. accuracy and  completion time.

%In each of 
%the figures, variables in different Tukey groups from the post-hoc analyses are shown in different colors horizontally. 
%For the interaction effect between encoding and  layout, we plot the performance of each encoding by layout.

%\begin{comment}

%\end{comment}

%\begin{comment}
%\end{comment}

\begin{figure}[t]
	\centering
	\begin{subfigure}{.5\linewidth}
		\centering
		\includegraphics[width=\linewidth ]{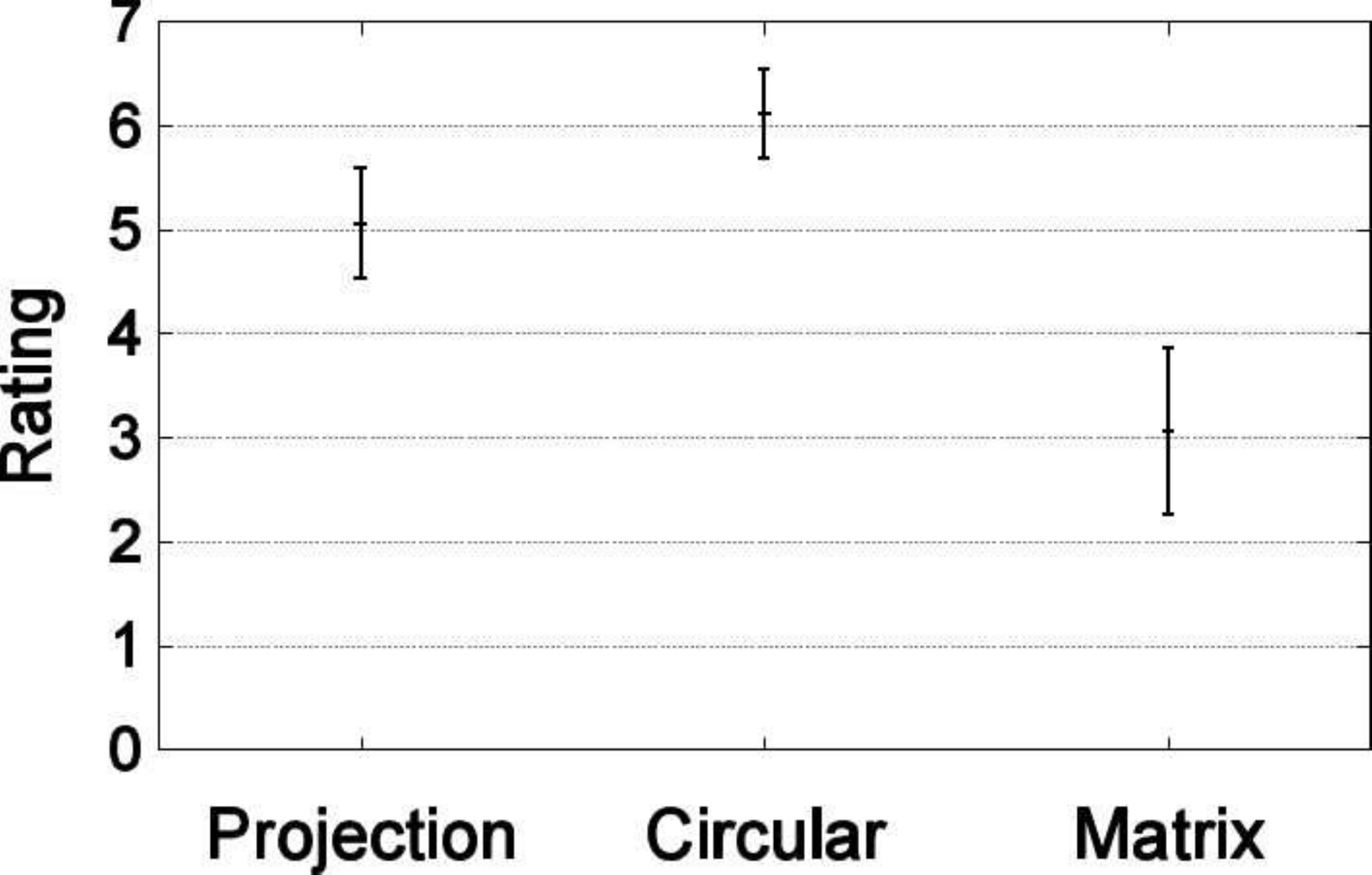}
		\caption{Layout vs. Rating}
		\label{fig:layRating}
	\end{subfigure}%
	\begin{subfigure}{.5\linewidth}
		\centering
		\includegraphics[width=\linewidth ]{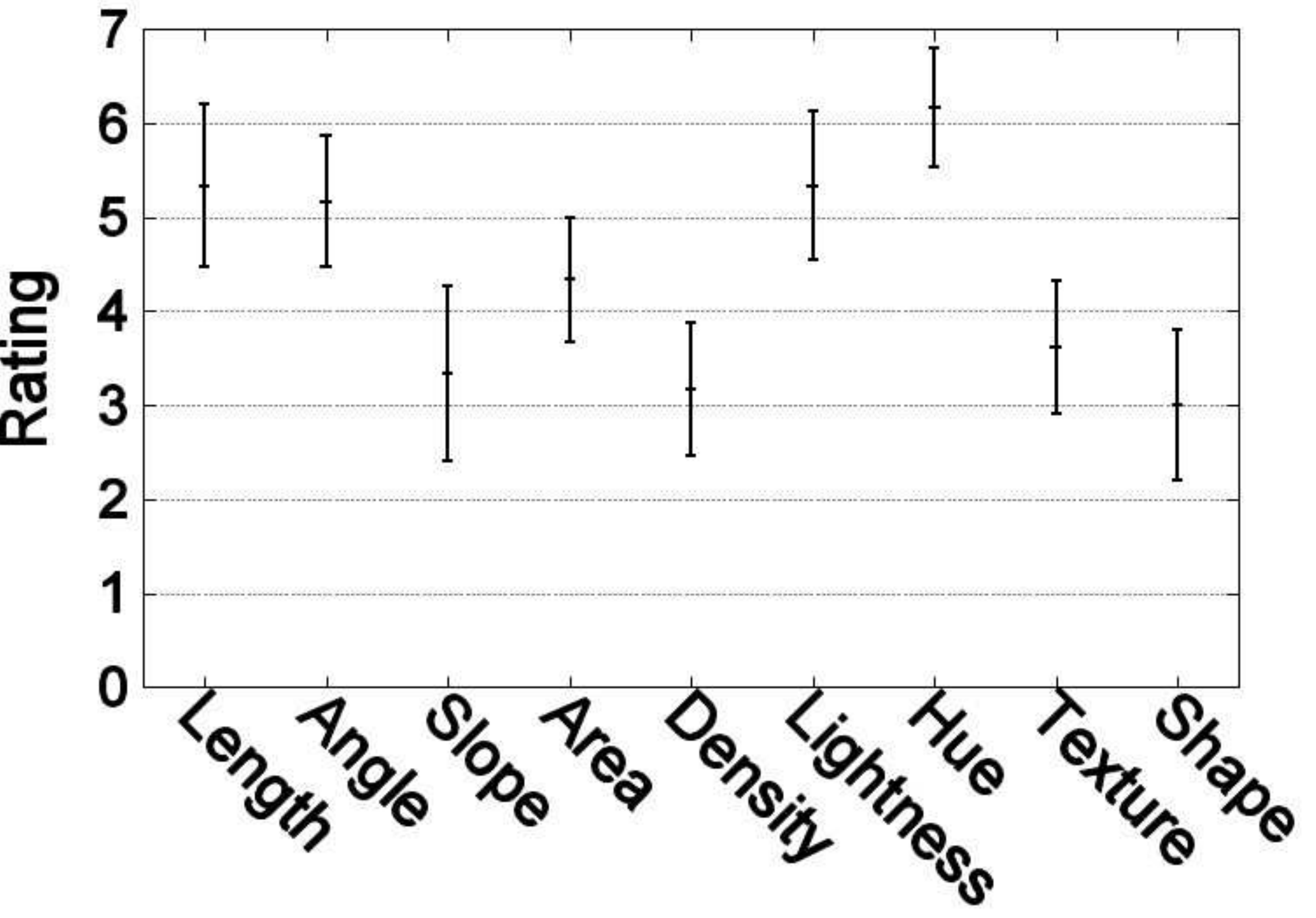}
		\caption{Visual Marker vs. Rating}
		\label{fig:encRating}
	\end{subfigure}%
	\caption{Post-questionnaire Subjective Ratings on Perceived Usefulness.}
	\label{fig:rating}
\end{figure}

\subsubsection{Task type 1: Change-detection Tasks}

This  task  asks  participants to compare which  is larger  between the same node  in two different networks placed side by side.  
We observe a significant main  effect  of layout on time (Fig.~\ref{fig:timeLayoutTask1}) 
and  the effect size is large  (d=0.40). The post-hoc  analysis shows   that  two methods of \textit{projection-matrix} and  
\textit{projection-circular}  belong   to  different  groups,  revealing  that \textit{projection} forms a group 
by itself and  \textit{circular}  and  \textit{matrix}
are in the same group.

We also  observe a significant main  effect of marker on correctness (Fig.~\ref{fig:accEncodingTask1}), 
though the  effect  size  of accuracy on encoding is small  ($V=0.15$). The marker groups revealed in post-hoc analysis are  
listed  in Fig.~\ref{fig:accEncodingTask1}. We  observe three  groups: lightness and  density (as expected) are in  the  least  
accurate group and  in  contrast hue-variation increases accuracy and falls into the same group as the most efficient 
texture, length, area, angle, and  shape. Slope can be in either  group.

%where colored dots along the same horizontal line are in separate group. 
\remove{It is worth noting that several layout-marker combinations (projection-length, 
projection-area, projection-hue, circular-angle, and circular-texture) lead 
to $100\%$ accuracy in this task type (Fig.{~\ref{fig:accIntTask1}}).}

%. Area, length, and  angle  are the most accurate 
%encoding; slope and lightness led to relatively poor accuracy compared to other  methods.

\subsubsection{Task type 2: NeighborHub Tasks}

This  task  asks  participants to locate  extremes in neighboring  nodes that  might be  spatially distributed in  different regions. 
Both layout and markers are significant main effects on both  accuracy and  completion time (Table~\ref{tab:expResults}). 
The effect sizes of the marker and  layout on time are both large.
%None of  the  two-way interactions was  significant. 

\textit{Circular} and  \textit{projection} are  most  accurate (Fig.~\ref{fig:accLayoutTask2}) and take least  task  completion 
time (Fig.~\ref{fig:timeLayoutTask2}). We  can  observe two  groups from  the  post-hoc analysis: 
\textit{projection}-\textit{circular} group and  \textit{matrix}  group.

Among the  markers, \textit{area}  and   \textit{angle}  lead  to  the  most accurate 
answers (Fig.~\ref{fig:accEncodingTask2}). \textit{Area}, \textit{length}, and  \textit{angle} 
take the least time for task completion (Fig.~\ref{fig:timeEncodingTask2}). \textit{Lightness} and  
\textit{slope}  are  the  least  accurate and  the  most time-consuming visual  markers.

%\textbf{Discussion.} 
%When we separate the encoding by layout, we see that ring-length and ring-size lead to perfect accuracy of $100\%$ and that ring-size had the shortest task completion time.

\subsubsection{Task type 3: LobeHub Tasks}

This  task  asks  participants to locate  the  extreme values of nodes within closely proximity.  
Both  main
effects  of  positioning and   marks   are  significant for  both completion time  and  correctness. 
The effect size is large  for time  but  not  accuracy. Here  \textit{circular}  and  \textit{matrix}  
have  the most  accurate answers (Fig.~\ref{fig:accLayoutTask3} and  
also take  the least time for task completion (Fig.~\ref{fig:timeLayoutTask3}). 
If we balance the accuracy and  time and  group techniques whenever possible,  the  post-hoc analysis reveals  two  
groups: the  \textit{matrix}-\textit{circular}  group and  the 
\textit{projection} group.
%The two-way  interaction of  positioning and  markers was  also  significant on  time (Fig.~\ref{fig:task3} (a)-(d)). 

The  effect  of visual  markers on  correctness is also  significant (Table~\ref{tab:expResults}), 
with  \textit{area},  \textit{texture}, and  \textit{angle}  being  the most accurate and \textit{slope} and 
\textit{density} being the least accurate. \textit{Area}  and  \textit{length}   
have  the  shortest task  completion  time;  \textit{slope}  extends task  completion time.  
The  post-hoc  analysis of accuracy 
reveals  three  larger  groups: \textit{area}, \textit{texture}, \textit{angle},  \textit{length},  and  \textit{hue} 
achieve  the most accurate answers,  \textit{density} and 
\textit{slope} the worst,  with \textit{shape}  and \textit{lightness} in between.

\remove{
When   the  accuracy and   task  completion time are separated by positioning techniques, the general trend was  that  projection reduced accuracy and  increased task completion time (Figs.{~\ref{fig:accIntTask3}},  {~\ref{fig:timeIntTask3}}). Marker effectiveness is similar  in  circular  and  matrix  views  except  in  two  orientation  markers, 
slope  and  angle.  
Density, lightness, and shape  are  markedly worse  for  projection than  for  circular and  matrix  positioning in terms  of correctness.
}

%\textbf{Discussion.} 

%The first characteristic may  explain why  we did  not  see accuracy advantages of circular  and matrix  views in task 2 since  the  nodes  are distributed in different positions along the circles or the straight lines.

\subsubsection{Task type 4: HemisphereHub Tasks}

This  task  asks  participants to  locate  symmetry and   find extremes in  neighbors belonging to  the  other  half  of  
the symmetrical network. We also  observe significant main  effects of both  layout and  encoding on correctness and  
time (Fig.~\ref{fig:task4} (a)-(d)). \textit{Circular} leads  to the most  accurate answers, followed by \textit{projection} 
and  \textit{matrix}.  
\textit{Circular} and  \textit{projection} lead to fast task completion. To balance  correctness and  accuracy, 
\textit{projection} and \textit{circular}  are 
in the same  group, leaving \textit{matrix}  a group by itself.
%Two-way interaction was also significant on time (Table~\ref{tab:expResults}). 
%All pairwise layout techniques fall into different Tukey  groups in  terms of correctness. 

Among the  visual  markers, \textit{density} surprisingly led  to the  most  accurate result,   followed by  the  \textit{hue}  and   
\textit{area}. 
The  other   group in  the  post-hoc  analysis  includes  all other  markers: 
\textit{length}, \textit{angle},  \textit{lightness}, \textit{shape}, \textit{slope},  and \textit{texture}.

%We also  observed that  accuracies using  density and  area were consistent across  three  positioning techniques. 
%Texture  and shape  influenced matrix  and  slope  and  lightness influenced projection more  than  any  other  techniques. 
%Matrix  led  to longer  task completion time regardless of visual  mark  selection in this task type.

\subsubsection{Subjective Rating}

Fig.{~\ref{fig:rating}} shows  the subjective ratings of  layout and  visual marker based  on  perceived 
usefulness, i.e., how  effective participants  think  certain  techniques could help  them  with  their  tasks  during the 
experiment. Overall, participants prefer  the \textit{circular}  layout, followed by \textit{projection}  and  then  \textit{matrix}.  
This result  
is mostly  in consistent with their objective performance results. Participants  rate  \textit{hue}  the highest, 
followed by \textit{lightness} among those  marker types. Participants do not think that  shapes are useful.

%Six subjects provided feedbacks after the post-questionnaire. Four of them commented on the visualization design. 
%Three of the four commented that layout using matrix was difficult to understand. Among them one suggested the instructions for matrix layout should be clearer and another commented that matrix visualization made tracing paths difficult and also made things cluttered. One subject thought hue encoding was good because the subject knew ``purplse was more than green reglardless of the nodes' postions'', but it was hard to compare slopes, lengths or areas when the nodes are far apart.

\section{Discussion} 

This   section   contains  the   design  knowledge  we   have gained from  the  ranking analysis and  hypothesis testing and  our explanation of the study results.

\subsection{Direct Encoding of Quantitative Values for Comparison Tasks}

Our   tasks   are   similar   to  that   of  Ghoniem,  Fekete,   and Castagliola~\cite{ghoniem2004comparison} and  
our results indicate that  visualizing quantitative variables as  much  as  possible increased task accuracy by 
$20\%$ in comparing quantitative values. While this  work  considers only one  dimension, a logical  next  step is  to  study  the   multivariate  data   analysis important  in genomics~\cite{gehlenborg2010visualization} and  connectomics~\cite{lichtman2014big}.

\subsection{Ranks of Visual Encoding}

\begin{figure}[!t]
	\centering
	\includegraphics[width=\linewidth ]{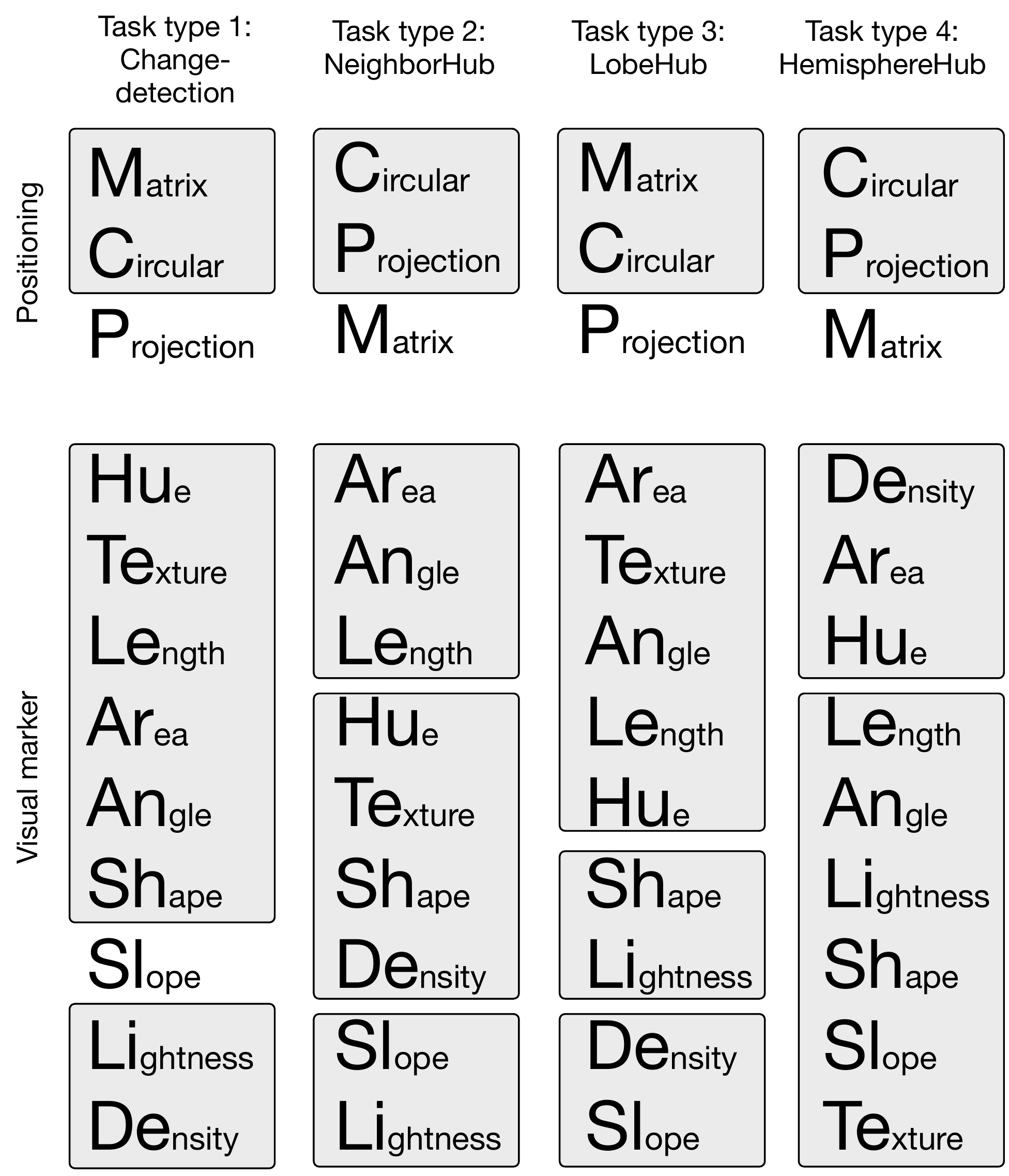}
	\caption{Visual  encoding ranks   by  comparison tasks.  The variables in the  same  gray  box belong  to the  same  group. The higher the box, the better  the variable performance.}
	\label{fig:ranking}
\end{figure}

Fig.~\ref{fig:ranking}  shows  the  variable ranking from  this  study.  Since one  of our  goals  is to rank  the  positioning and  marker encodings,  we  have  balanced   accuracy  and   time   and group techniques together when handling layout; To the visual
marker ranking, 
we have followed Mackinlay~\cite{mackinlay1986automating} by accuracy only. 
We group two markers if they are not  in significantly different groups in the  
post-hoc analysis reported in Figs.~\ref{fig:task1}-\ref{fig:task4}.

Our results in general are also consistent with  those of Garlandini and  Fabrikant's work~\cite{garlandini2009evaluating}  
in  terms  of rank order of  \textit{size}, \textit{hue},  and \textit{orientation} for change detection. Their work found 
that size was most and orientation was least accurate and  hue  was  most visually salient. 
Our  variable space is larger  than  theirs.  Our rank  of \textit{hue} is much higher
perhaps because our multiple-hue map supports clear differentiation of 
close numerical values~\cite{ware1988color}.

We can  make  several  interesting observations. For the positioning techniques, \textit{circular} is always among the  best; 
\textit{matrix}  is superior for  tasks that  do  not  require \textit{symmetry} detection (e.g.  for  Change-detection and  LobeHub) 
or when nodes are close adjacent to each other  (e.g., nodes in LobeHub are  closer  than NeighborHub).

Among the nine visual markers, \textit{hue} and \textit{area} are the two markers that are always in the  top  groups;
\textit{length}  is also 
except for the final complex HemisphereHub task type. This result is considered to be significant, 
especially considering that \textit{color}
is the most used quantitative data encoding in the brain science domain~\cite{christen2013colorful}: our results
demonstrate that a carefully chosen color map can be effective to address complex data visualization.  An immediate future 
direction is to study what color maps would be more  effective for a broader range  of comparison tasks.  
Since \textit{lightness} in general is less accurate despite its monotonic luminance, 
our results show that adding \textit{hue} variations would be more effective in showing quantitative values
and our adoption of the multi-hue clearly  
shows that  hue  supports discriminations at various scales. 
%when the  size  markers  are  not  always spatially aligned.

Another  interesting  result   relates  to  \textit{shape}: \textit{shape} appears to  be  good  for  simple   
comparison (e.g.,  change-detection) but  poor for many comparisons. 
This result might show  that \textit{shape} with curvature is difficult to scale, 
even though two-curvature comparison is among the easiest.

\textit{Texture}  is suitable for local closely adjacent comparisons (e.g.,  Change-detection and  LobeHub) but 
when distances between nodes  increase, \textit{texture}  does 
not   fit  well   (e.g., neighborHub and  HemisphereHub). \textit{Slope},  \textit{lightness}, and \textit{density} are the three  
visual  markers frequently ranked low which is in consistent with literature.
It is unclear why \textit{density} is appeared to be the best for complex HemisphereHub tasks.

\subsection{Hypothesis Testing}
%\subsection{Visual Layout Effectiveness and Efficiency}

%\begin{comment}
\noindent\fbox{
	\parbox[l]{0.47\textwidth}{
\textit{For change-detection tasks, we would not observe differences in correctness among the three layout approaches. [Supported]} 
	}}
	\vspace{1mm}

H1 is supported (Fig.~\ref{fig:accLayoutTask1}). It is not  surprising that  these three  positioning techniques produce  about   
the  same  accuracy,   since  we  believe   the  result   is  influenced mainly by the  spatial proximity of the  two  values compared, 
and the proximities of all side by side comparisons in the three positioning techniques are  the  same.  \textit{Matrix}  
takes   longer and  \textit{projection} is most  efficient (Fig.~\ref{fig:timeLayoutTask1}), 
perhaps because participants  tend  to  get  an  overview of the  entire  layout before answering the questions. 
\textit{Projection} is most  straightforward in layout, followed by  \textit{circular}  and  \textit{matrix}  views,  
as remarked by our brain scientist collaborators.
%look at  the node and then trace to the visual marks. 

%\subsection{Visual Marker Effectiveness and Efficiency}

%\subsection{Layout for Visual Search}
	\vspace{2mm}

\noindent\fbox{
	\parbox[l]{0.47\textwidth}{
		\textit{
H2. Circular  and matrix  would be more accurate than  projection  in  the  LobeHub tasks,  but  circular  and  projection would  be more accurate  than  matrix  in  NeighborHub  and HemisphereHub tasks. [Supported]
		}
	}}
	\vspace{1mm}

This  hypothesis is supported. One  way  to  explain  the efficiency of \textit{circular}  and \textit{matrix} views for local region search
tasks,  such  as the  LobeHub tasks,  is that  the  ordering and alignment of the nodes on these two methods 
improve accuracy. These  tasks  also  do  not  need  symmetrical search, which would otherwise work  badly for 
the \textit{matrix}.  
\textit{Circular} may have the most  ordered linear  arrangement of nodes in a lobe which might explain its lowest  task completion time.
 
%Matrix,   which uses a  jigsaw
%pattern  of   node   marker  arrangement,  has   the   second best completion time, while  projection has on average the longest completion  time.

The order of the layout efficiency changed for the NeighborHub and  the HemisphereHub tasks, where the 
\textit{projection} and  \textit{circular}  lead  to more  accurate answers than  the \textit{matrix}. 
For all these three  tasks,  participants need  to conduct two subtasks: (1)  visual   search   of  those   nodes  within  a lobe  (for  lobeHub), or  a  region   (for  neighborHub), or  a hemisphere (for  hemisphereHub) 
and  (2) comparisons  of centrality measures on these  nodes.
 %Since these nodes can be highlighted, comparisons made after highlighting
%LobeHub tasks were benefited from the common spatial references either on  a straight line  or on  
%a circle for the matrix  and  circular  projections. Also, nodes  in the lobes are  spatially adjacent to each  
%other in the  matrix and circular views. 
The result  that comparison becomes worse  for \textit{matrix} than 
\textit{circular}   and   \textit{projection}  on  NeighborHub and   HemisphereHub tasks can be explained by the two 
different types of structural symmetry: orthogonal for \textit{matrix}  and  \textit{mirror} for the  other  two  methods. Here  
orthogonal symmetry on two perpendicular axes in matrix  increases task completion time.  That \textit{matrix}  increases 
task time can also be explained by the lack of patterns in neighboring node search,  causing difficulty in finding neighboring nodes.
%Type of symmetry of node positioning techniques 
%supporting the symmetrical scan (here, circular  and projection) 
%led to higher accuracy than  the  technique that  does  not  support symmetry  (here matrix) for 

That \textit{projection}  is  always  worse   than  \textit{circular}   can  be perhaps explained by  its  lack  of  boundary which   
makes interactive visual search  of  neighbors harder.  Though it is  also  generally  believed that  our 
boundary-supporting Gestalt  principle of ``closure'' facilitates grouping, that humans  use  convex  hulls  to  
enforce  closure~\cite{van2008perceptual}, and
that proximity also  implicates groups and  similarity~\cite{fabrikant2005cognitively}, 
none of these  benefited \textit{projection}, 
as was true  for the three  medical professionals as  well.  We  may  want  to  use \textit{projections} with caution, 
as in Jianu et al.~\cite{jianu2012exploring}, for brain image  comparison tasks.

\begin{comment}
These results confirmed three design principles: (1) spatial layout or positioning is a strong cue for interpreting 
quantitative values in network visualization; (2)
For tasks requiring users to make use of symmetrical pattern, use the mirror symmetry in the circular and projections instead of
the orthogonal symmetry in matrix view. 
%symmetrical layout of circular and projection would influence task
% influences effectiveness and that layout that directly addresses task needs  will  lead  to the  most  accurate answers;
And (3) circular layout perhaps supports symmetry, ordering, and spatial proximity thus led to the most
accurate answers in nearly all task conditions. 
\end{comment}

	\vspace{2mm}

\noindent\fbox{
\parbox[l]{0.47\textwidth}{
\textit{For Change-detection tasks, the ranking of the visual markers  partially follow the Mackinlay order for quantitative  data of length, angle, slope, area, density, lightness, and hue. [Partially supported]
 } 
	}}
	\vspace{1mm}

For mark effectiveness, the general rank order of selected marker types  follow  the original Mackinlay order reasonably well  
only for  length, area,  and  lightness. Here  hue showed  greater  competency and   fell  in  the  same   group as  the  most  
effective  length. One  explanation is the  color marker choices -- multiple hues   are  effective   for  finding large  and   
small~\cite{ware1988color}.  
The  curvatures used   in  the  shapes are  not  as  pre-attentive,  especially for  complex  tasks involving many  
item comparisons, but can be effective for  two  items; this shows   its  scalability limitations,  as  we expected, perhaps because of participant lack of familiarity with shape  markers.

	\vspace{2mm}

\noindent\fbox{
	\parbox[l]{0.47\textwidth}{
		\textit{
H4. For multi-scale comparisons, the rankings for hue, texture, and shape would improve, though 
they are ranked as the last  three in  the Mackinlay order for showing quantities. [Partially supported]}
		}
	}
	
	\vspace{1mm}

This hypothesis was  partially supported: ranking improved for \textit{hue} and \textit{texture} but not \textit{shape}. 
\textit{Texture}   led  to  the  most   accurate  answers,  especially for the   change-detection and   LobeHub tasks,
perhaps because this band texture allows  both size and  frequency detection. 
%Shape,  as  discussed, makes  use  of curvature but is not as scalable  as we would expect.
%and length had about the same accuracy suggesting that the use of pre-attentive cues
%could greatly improve accuracy. 
%The lack of dramatic improved performance of shape over its previous ranking may 
%be due  to the 
%different definition of shapes. In our experiment, we actually change  the  curvature of the  marker boundary thus 
%making use of
%pre-attentive attributes to order the shapes.

There  is great  interest in the  community in using  \textit{shape} or glyphs in design. 
Though we have  adopted ordered glyphs, the  
accuracy of this ordered glyph and  task-completion time are not as good as we had hoped. Our result  would 
suggest that  single-variable encoding with  \textit{shape} 
has  limited use, and  yet multiple encodings of \textit{shapes} (e.g., tensor  encodings in 2D brain  imaging visualization 
in Laidlaw et al.~\cite{laidlaw1998visualizing} and tensor glyphs 
in Schultz and Kindlmann~\cite{schultz2010superquadric}) and Kindlmann~\cite{kindlmann2004superquadric} 
need to be studied
since  they  make possible  new relationship tasks  for which  design recommendations  are 
just  appearing~\cite{maguire2012taxonomy}.  
\textit{Hue} also improves performance and  leads to about  the  same  accuracy  as area, suggesting that combing 
multi-hues and luminance convey order effectively.

	\vspace{2mm}

\noindent\fbox{
	\parbox[l]{0.47\textwidth}{
		\textit{				
H5. Length in general would lead to the most accurate answers and  would consume the least amount  of time. [Partially  supported] 
		}
	}}
	\vspace{1mm}
	
%Fig. ~\ref{fig:accIntTask3} and Fig.~\ref{fig:timeIntTask3} we see that  size encoding with  ring  layout leads  to  the  best  performance in  terms of accuracy and  completion time. It is better  than  length encoding with  matrix  by a small margin. Though both combination has the same  accuracy (100\% in this case), size encoding with  ring ranks  higher because  it leads  to slightly faster  task completion.

Our results are contrary to those  in Cleveland  and  McGill~\cite{cleveland1985graphical}. Length  was  better  in  
Cleveland and McGill study while  area and angle were more accurate than length  for the relatively complex tasks  
in task  types  2-4. This has to do with  the limited number of distinguishable steps. Cleveland and  McGill's experiments 
on reading charts  drill down to very  small  differences in encoded value,  while  in our  experiments, our  task-generation 
criteria   ensure that the  difference is 
relatively large  ($5\%$); this  eliminates cases with  very  small  visual  difference, where the aligned length encoding 
should be superior.

%It is unclear why this occurs and more controlled studies will be needed to explain this result. 

%For tasks  that  require comparing values  of nodes  that can  be arranged close to each  other,  we  recommend using matrix  
%layout combined with  length encoding. Size encod- ing  is also  recommended for  this  task  and  performs well with  
%all three  layout methods.
%It is perhaps not surprising to learn area and length are better as they are the highly ranked markers in the 
%Mackinaly study for quantitative tasks. Here we show that they (at least area) are also effective 
%for comparisons tasks at various scales.

\subsection{Quantitative Overlaying on Topological Networks}

%Though many   visualization studies have  been  conducted on representing the syntactic information of networks, semantics encoding studies .  
%Unlike research on syntactic information encoding where layout algorithms and edge bundle techniques 
%are the end products, 
 
Our  study fills in the literature by looking at node  attribute encoding for network visualization. Our major  take-away 
messages  are  that  this  explicit  encoding improves accuracy by at least $20\%$ compared to the former  empirical study results and ranking orders for comparison tasks at various scales of node  distributions. These  results can assist  
visualization  designers to choose  quantitative encoding methods for network comparison tasks.

We focus on evaluating the effectiveness of several  visual markers in conveying node  attributes on three  layout conditions. 
A somewhat surprising result  was  that  \textit{hues}  in various scale conditions works  well except  in searching  among a large  
group of neighboring nodes. In addition to  brain  network visualization, many  other  networks such as biological  
pathway  visualizations also  require encoding quantitative values.  Brain-image-based visualization 
tools  (such  as  AFNI~\cite{cox2012afni}, BrainVoyager~\cite{goebel2012brainvoyager}, and  LONI~\cite{rex2003loni}) 
rely  heavily on  color  to represent statistics of brain regions. Network-based visualization 
tools (such  as BrainNet Viewer~\cite{xia2013brainnet}
and  Connectome Viewer~\cite{gerhard2011connectome})  
use  color  and  size  pervasively to  encode node   attributes. 
Our study has shown that  visual  variables other  than  \textit{hue} and \textit{area}  (as well as carefully designed texture)  
can  in many  cases  achieve  similar  if not  better performance. For network layout, we  note  that  though the  brain  research community has  adopted some  layout techniques to  show brain  region  clusters~\cite{nelson2010parcellation}, 
inefficient  text  labeling or  additional  views  are required to associate nodes with  individual brain  regions.

%Our results are complementary to those of Alper et al.~\cite{alper2013weighted} in that they focus on edge attributes while we focus on node attributes with  a large  selection  of visual  variables. 
%This choice of focus gives our  evaluation results more  applicability for   visualization designers  faced  with  choosing encodings for node  attributes in brain  network visualizations.
%Few techniques enable  syntactic information to be represented within a spatial context.  Our research has  just  begun looking at these issues.

\subsection{Strength and Limitations}

The strength of the  present experiment is its inclusion of a broad range  of visual  variables and  several  commonly used 
layout methods. Nine  markers and  three  layout techniques are carefully compared. The reliability of the experiment is 
enhanced by  the  well-controlled task difficulty. Difficulty is controlled according to  each task and the  
underlying data characteristics. This reduces the  influence  of data  variation on human performance and  makes  it more  likely that 
the performance variability we see actually comes from  a variation in effectiveness among the different visual markers and 
layout methods. Our experiments have limitations: like any  lab study involving human participants, our study results may 
be different if we use medical professionals.

%We select three  layout techniques from  existing  brain  research publications for generic  graph visualization, making the results gained from  the experiment easily  applicable for brain  functional network visualizations.
%relatively small  and  may  not well represent a larger  population. 
%Future work could involve  more   participants  and   even  use  crowd-sourcing to generate a larger  sample. 
%Second,  a generic  ranking of encoding or layout for general usage cannot be reached and will always depends on task types. 
%Giving the number of  visual  variables involved  in  this  experiment, larger  sample sizes might also be required to generate a 
%robust ranking, even for a very  specific task. Future study could  focus  on  a  smaller set  of visual variables 
%based  on the results of this experiment.

\section{Conclusion}

Our study evaluates network visualization by selecting and comparing nine  visual  markers and  three  layout methods. Our  experimental results provide the following recommendations for designing single-variate network visualizations.

%Tasks that put more emphasis on the spatial information of nodes  than previous network
%visualization research are  included in our  user  and  the  task  difficulty is carefully controlled. 

%In this study, we implement visual markers for showing node centrality in brain fMRI network visualization using nine visual variables. We select a subset form our task topology and conduct user study to evaluate the effectiveness of nine types of markers and three layout schemes. Our results indicates a different ranking from previous work due to controlled minimal difference among nodes. We also discover that a linear arrangement of nodes could greatly facilitate comparing node values within a group.

\begin{itemize}
\item 
Positioning techniques influence task  effectiveness; positioning that  is closely  proximate to  task  conditions is likely to achieve  the best performance.

	\begin{itemize}
		\item When  tasks  are  related to symmetrical structure,  use  circular  or projection symmetrical positioning.
		\item When tasks are related to reading from a set of randomly distributed values in closely 
		proximate spatial locations (brain  lobes), use matrix or circular  positioning.
	\end{itemize}
	
\item For  pairwise comparison tasks,  monotonic luminance multi-hue, texture, length, area, angle,  and  curvature-inspired shape achieve the best results.

\item For  multiple comparisons in  closer  adjacent areas,  area,  texture, angle,  length, and  hues  are the best.

\item Avoid  lightness and  slope  as much  as possible for data comparison.
% for group value comparisons.

%\item Matrix  layout combined with  length encoding is best  for  comparing values  of nodes  arranged together.

\end{itemize}

\appendices
\section{Pilot Study Results: Comparison of Two Background Coloring Schemes}

\begin{figure*}[t]
	\centering
	\includegraphics[width=\linewidth]{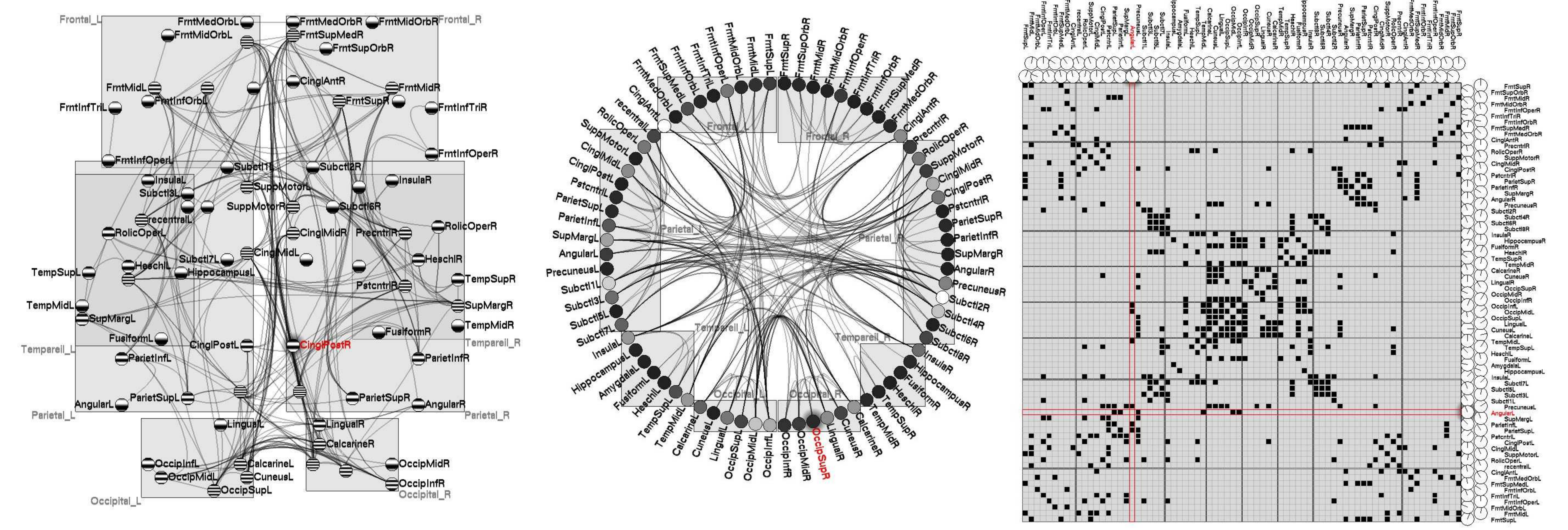}
	\caption{The white-background color scheme.}
	\label{fig:whiteBackground}
\end{figure*}

We have compared the white and the gray background color schemes to understand the effect of background on 
task completion time and accuracy (i.e., percentage of correct answers.) 
The gray-background coloring scheme uses $50\%$ gray as the background color and  
outline the markers in black to increase contrast. The white-background networks depict the 
figures in black (Fig.~\ref{fig:whiteBackground}). This study follows the same procedure as the formal study
except that we only tested the sparse graph and half of the participants did black and half white first. Data assignment
was randomized. This pilot study showed that the background was not a significant main effect either on task completion time or on 
accuracy using the same statistical measurement methods as the formal study (Table~\ref{tab:expResultsBackground}).

\begin{table}[!th]
%% Table captions on top in journal version
 \caption{Summary Statistics by Tasks (The Background Effect Pilot Study).}
 \label{tab:expResultsBackground}
 \scriptsize
 \centering
   \begin{tabular}{c | l | l  }
   \hline
     Task & Variable & Statistical test \\
   \hline 
                &  Accuracy (marker)           &  $\chi^2_{(8, 324)}$=5.72, p=0.7  \\
                &  Accuracy (layout)            &  $\chi^2_{(2, 324)}$=2.75, p=0.25  \\
Change-   & Accuracy (background)   &  $\chi^2_{(1, 324)}$=1.40,  p=0.24 \\ 
detection  &  Time (marker)                &   $F_{(8, 324)}=1.0$, p=0.41  \\
                 & Time (layout)                  &   \textbf{$F_{(2, 324)}=16.45$}, p$<$0.0001   \\ 
                & Time (background)          &     $F_{(2, 324)}=0.22$, p=0.64   \\
         %   &  \textbf{Time}         &  \textbf{F(16, 948)=1.8}               & \textbf{d=1.15}\\ 
          %   & (\textbf{marker $\times$ layout}) & \textbf{p=0.03} & \\
    \hline
  
                    & Accuracy (marker)        &  $\chi^2_{(8, 324)}=10.2$, p=0.25                 \\ 
                    & Accuracy (layout)    & $\chi^2_{(2, 324)}=3.9$, p=0.14               \\ 
   Neighbor   & Accuracy (background) &  $\chi^2_{(2, 324)}=1.32$,   p=0.25 \\
          Hub    & Time (marker)       & $F_{(8, 324)}=1.87$, p=0.065        \\
                     & Time (layout)          & \textbf{$F_{(2, 324)}=74.32$, p$<$0.0001}             \\ 
                     & Time (background)        & $F_{(2, 324)}=0.59$, p=0.44          \\ 
         %      & \textbf{Time} \textcolor{red}{NEW}     & \textbf{F(16, 950)=2.0} & \textbf{d=1.89}\\
          %      & \textbf{(marker $\times$ layout)} & \textbf{p=0.01} & \\ 
               \hline
               
                      &  Accuracy (marker)       & $\chi^2_{(2, 324)}=3.92$, p=0.86              \\ 
                      & Accuracy (layout)            & $\chi^2_{(2, 324)}=2.30$, p=0.32             \\
      Lobe        & Accuracy (background)            & $\chi^2_{(2, 324)}=0.065$, p=0.8    \\
         Hub       &  Time (marker)  & \textbf{$F_{(8, 324)}=6.97$, p$<$0.0001}   	         \\
                       &  Time (layout)       &  $F_{(2, 324)}=2.49$, p=0.08                \\ 
                       &  Time (background)       &  $F_{(2, 324)}=1.52$, p=0.22    \\ 
       %       &  \textbf{Time}         &  \textbf{F(16, 952)=2.1}              & \textbf{d=1.86}\\
        %       & \textbf{(marker $\times$ layout)} & \textbf{p=0.009} & \\ 
    \hline 
    
                         &  Accuracy (marker)       & $\chi^2_{(2, 324)}=8.14$, p=0.42              \\  
                         &  Accuracy (layout)          &  $\chi^2_{(8, 324)}=1.71$, p=0.43   \\
  Hemisphere	&  Accuracy (background)         & $\chi^2_{(8, 324)}=0.1$, p=0.90   \\
  	 Hub         &  Time (marker)  & $F_{(8, 324)}=0.7$, p=0.68   \\
                         &  Time (layout)   & \textbf{$F_{(2, 324)}=110.7$, p$<$0.0001}             \\ 
                         &  Time (background)   & $F_{(1, 324)}=0.02$, p=0.89             \\ 
      %     &  \textbf{Time}         &  \textbf{F(16, 952)=2.5}             &\textbf{d=2.02}\\ 
      %        & \textbf{(marker $\times$ layout)} & \textbf{p=0.001} & \\
               \hline   
    \end{tabular}
\end{table}

%\subsection{}

\ifCLASSOPTIONcompsoc
  \section*{Acknowledgments}
\else
  \section*{Acknowledgment}
\fi

The authors would like  to thank Drs.  Susumu Mori,  Judd Storrs,  and  Kenneth L. Weiss for their  input on the task  selection. 
The authors would also like to thank the participants for  their  contributions and  the  anonymous reviewers for their  insights. 
Special  thanks to Katrina Avery  for revising the  manuscript. This  work  was  supported in  part  by  
National  Science Foundation (NSF) awards IIS-1302755, CNS-1531491, ABI-1260795, EPS-0903234, and  DBI-1062057. Any  opinions, findings, and  conclusions or  recommendations expressed in this material are those  of the authors and do not necessarily reflect  the views  of the National Science Foundation.

Jian Chen is the corresponding author.

\ifCLASSOPTIONcaptionsoff
  \newpage
\fi


\bibliographystyle{IEEEtran}
\bibliography{fMRIEval}
%\begin{thebibliography}{1}


%\end{thebibliography}


%\end{IEEEbiography}


%\newpage
\begin{IEEEbiography}
	[{\includegraphics[width=1in,height=1.25in,clip,keepaspectratio]{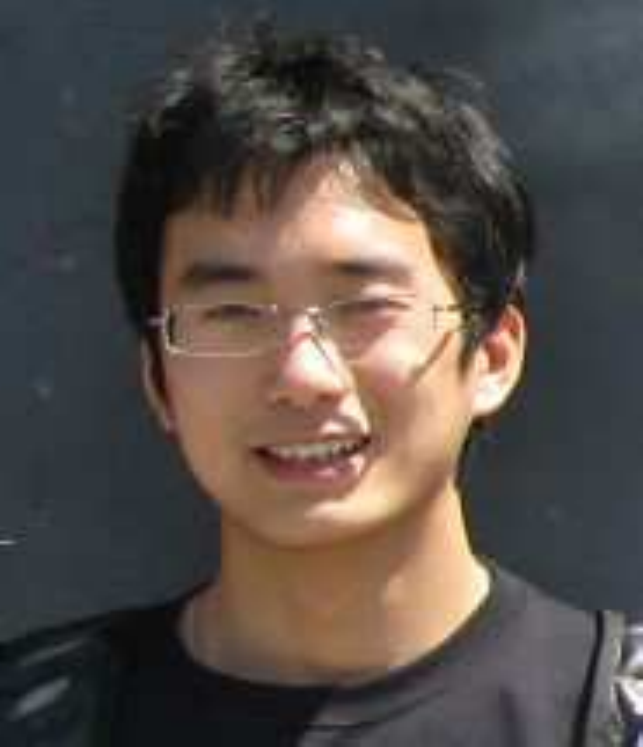}}]
	{Guohao Zhang} is a PhD student at University of Maryland, Baltimore County. He received his B.E. degree in Engineering Physics from Tsinghua University in 2012. His research interests include design and evaluation of visualization techniques and 3D visualizations. He is a student member of IEEE.
\end{IEEEbiography}

\begin{IEEEbiography}
	[{\includegraphics[width=1in,height=1.25in,clip,keepaspectratio]{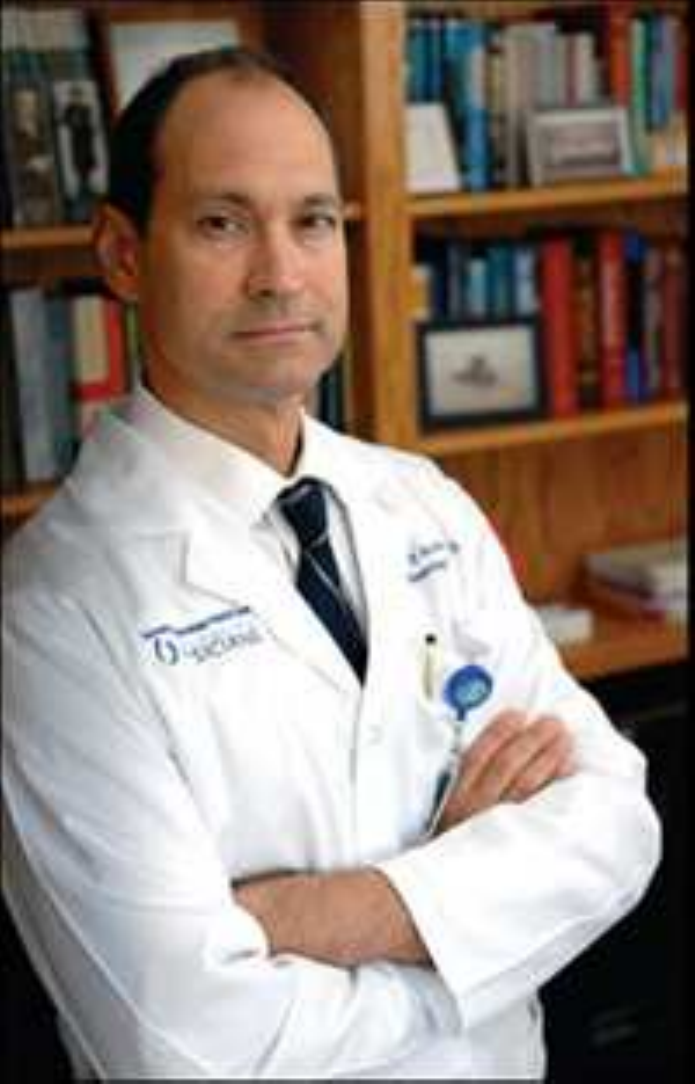}}]
	{Alexander P. Auchus}
	Dr. Alexander P. Auchus holds degrees from Johns Hopkins University and from Washington University in St.Louis.  He is an elected fellow of the American Neurological Association, the American Academy of Neurology, and the American Geriatrics Society.  He has served on the faculty of Emory University, Case Western Reserve University and University of Tennessee.  His present position is Professor and McCarty Chair of Neurology at the University of Mississippi Medical Center.  Dr. Auchus'??s research interests are in neuroimaging biomarkers for Alzheimer's disease and other dementias.
\end{IEEEbiography}

\begin{IEEEbiography}
	[{\includegraphics[width=1in,height=1.25in,clip,keepaspectratio]{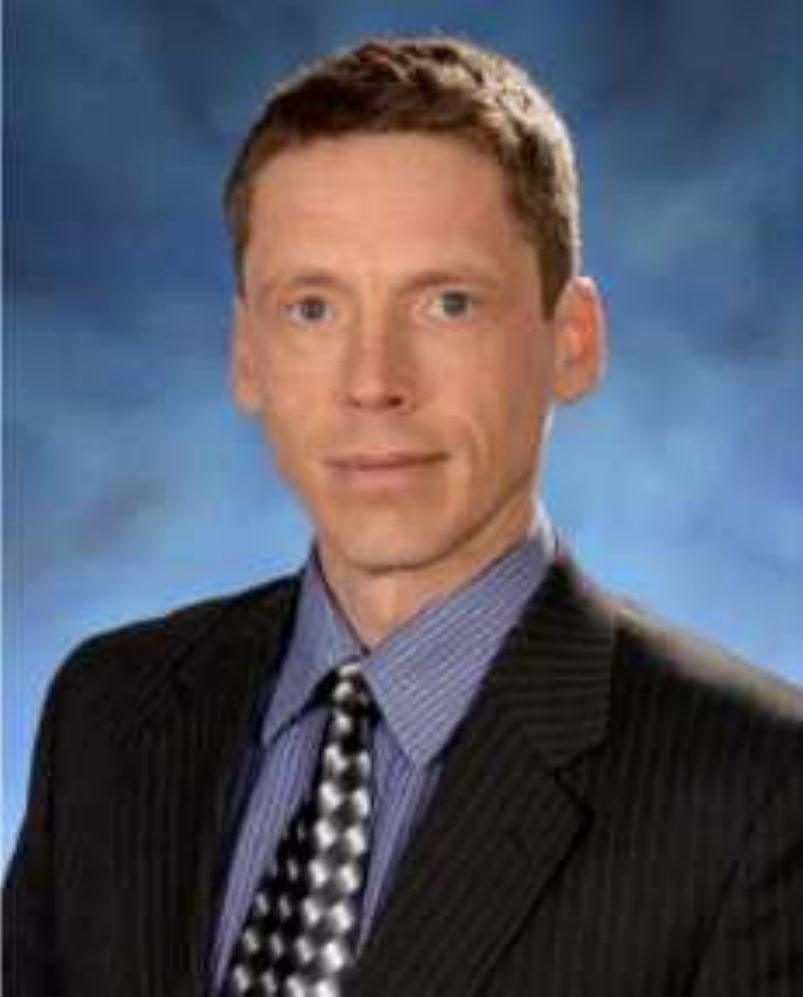}}]
	{Peter Kochunov}
	 Dr. Peter Kochunov is a professor in the Maryland Psychiatric Research Center and an active
	 collaborator on projects within the Neurobehavioral Research Division. Dr. Kochunov is also Head of 
	 the Human MRI group for the RII and a staff scientist at the Southwest 
	 Foundation for Biomedical Research. He is a board certified medical physicist with an active research
	 program in genetic imaging. ??Besides his research, Dr. Kochunov has directed a variety of graduate courses
	 on medical imaging and fMRI techniques. He also is the developer of the BrainVisa Morphological extensions.
\end{IEEEbiography}

\begin{IEEEbiography}
	[{\includegraphics[width=1in,height=1.25in,clip,keepaspectratio]{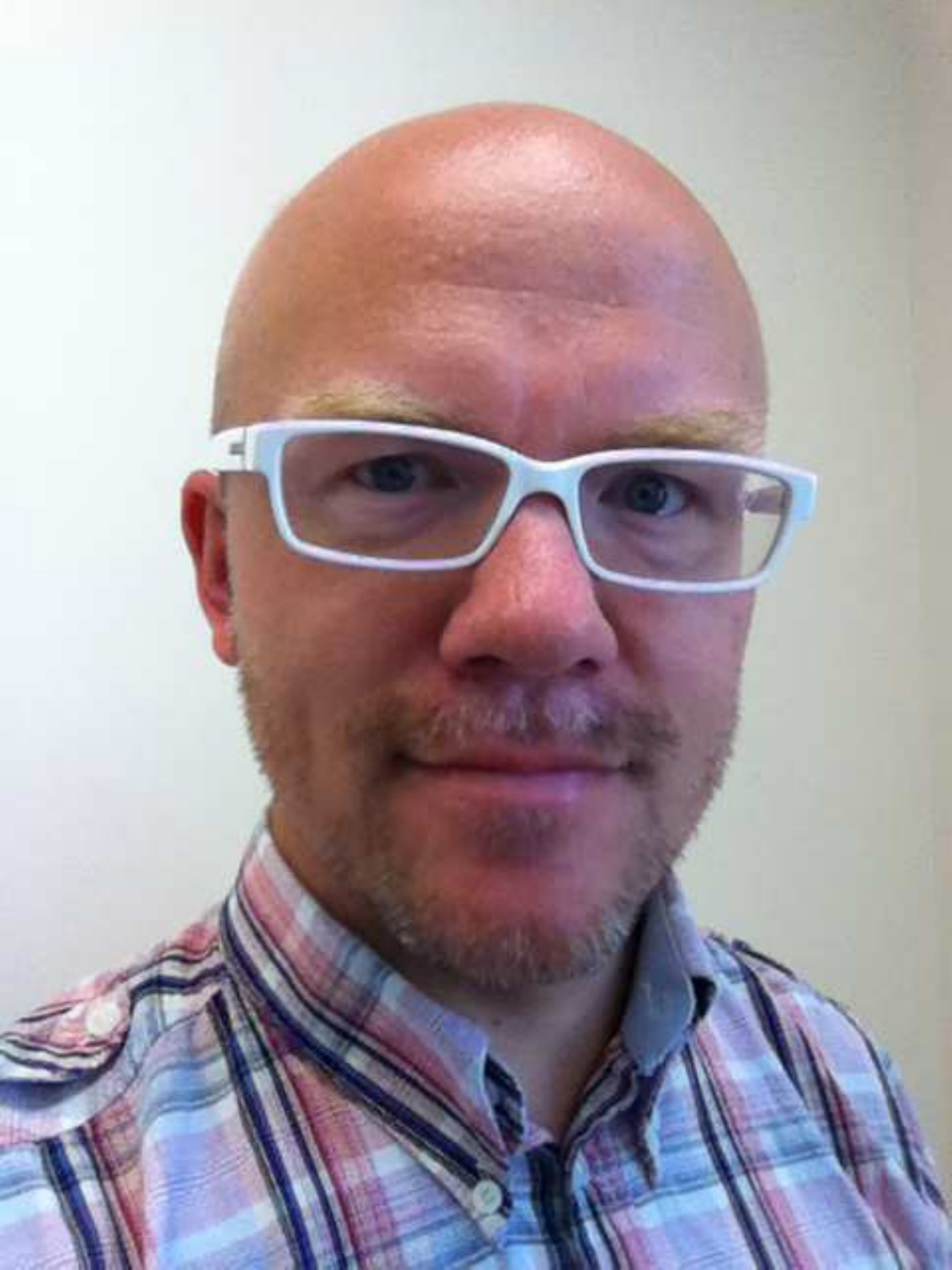}}]
	{Niklas Elmqvist}
	 Dr. Niklas Elmqvist received the Ph.D. degree in 2006 from Chalmers University of Technology in
G\''{o}teborg, Sweden. He is an associate professor in the College of Information Studies at University
of Maryland, College Park, MD, USA. He was previously an assistant professor in the School of Electrical \& Computer Engineering at Purdue University in West Lafayette, IN. He is a senior member
of the IEEE and the IEEE Computer Society.
\end{IEEEbiography}

\begin{IEEEbiography}
	[{\includegraphics[width=1in,height=1.25in,clip,keepaspectratio]{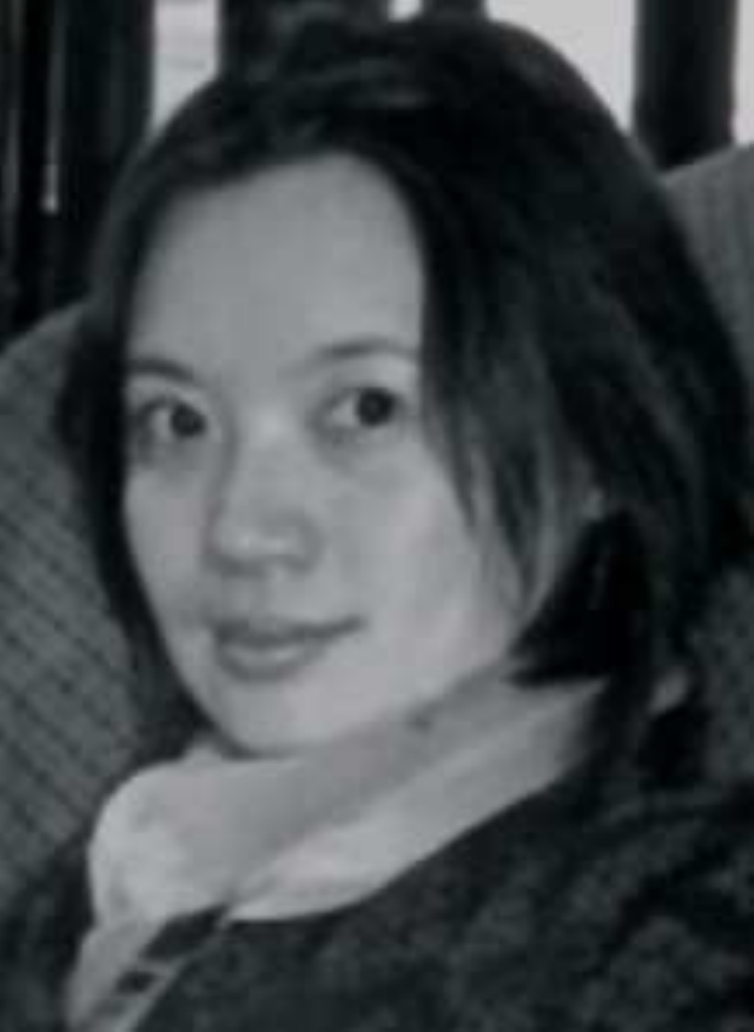}}]
	{Jian Chen} Dr. Jian Chen received the PhD degree in Computer Science from Virginia Polytechnic Institute and State University (Virginia Tech). She did her postdoctoral work in Computer Science and BioMed at Brown University. She is an Assistant Professor in
the Department of Computer Science and Engineering at The Ohio State University, where she directs the 
Human-in-the-Loop Visual Computing Lab. She is also affiliated with the Ohio Translational Data Analytics Institute. 
Her research interests include design and evaluation of visualization techniques, 3D interface, and immersive analytics. 
She is a member of the IEEE and the IEEE Computer Society.
\end{IEEEbiography}




%\vfill


\end{document}